\documentclass[twocolumn,showpacs,preprintnumbers,amsmath,amssymb,mathabs,nath,floatfix]{revtex4}

\usepackage{graphicx}
\graphicspath{{./figs/}}
\usepackage{dcolumn}
\usepackage{bm}
\usepackage{subfigure}

\begin{document}


\title{An Improved Limit on the Muon Electric Dipole Moment}

\author{
G.W.~Bennett$^{2}$, B.~Bousquet$^{10}$, H.N.~Brown$^2$,
G.~Bunce$^2$, R.M.~Carey$^1$, P.~Cushman$^{10}$, G.T.~Danby$^2$,
P.T.~Debevec$^8$, M.~Deile$^{13}$, H.~Deng$^{13}$,
W.~Deninger$^8$, S.K.~Dhawan$^{13}$, V.P.~Druzhinin$^3$,
L.~Duong$^{10}$, E.~Efstathiadis$^1$, F.J.M.~Farley$^{13}$,
G.V.~Fedotovich$^3$, S.~Giron$^{10}$, F.E.~Gray$^8$,
D.~Grigoriev$^3$, M.~Grosse-Perdekamp$^{13}$, A.~Grossmann$^7$,
M.F.~Hare$^1$, D.W.~Hertzog$^8$, X.~Huang$^1$,
V.W.~Hughes$^{13,\dagger}$, M.~Iwasaki$^{12}$,
K.~Jungmann$^{6}$, D.~Kawall$^{13}$, M.~Kawamura$^{12}$,
B.I.~Khazin$^3$, J.~Kindem$^{10}$, F.~Krienen$^{1,\dagger}$,
I.~Kronkvist$^{10}$, A.~Lam$^1$, R.~Larsen$^2$, Y.Y.~Lee$^2$,
I.~Logashenko$^{1,3}$, R.~McNabb$^{10,8}$, W.~Meng$^2$, J.~Mi$^2$,
J.P.~Miller$^1$, Y. Mizumachi$^{11}$, W.M.~Morse$^2$,
D.~Nikas$^2$, C.J.G.~Onderwater$^{8,6}$, Y.~Orlov$^4$,
C.S.~\"{O}zben$^{2,8}$, J.M.~Paley$^1$, Q.~Peng$^1$,
C.C.~Polly$^8$, J.~Pretz$^{13}$, R.~Prigl$^{2}$,
G.~zu~Putlitz$^7$, T.~Qian$^{10}$, S.I.~Redin$^{3,13}$,
O.~Rind$^1$, B.L.~Roberts$^1$, N.~Ryskulov$^3$, S.~Sedykh$^8$,
Y.K.~Semertzidis$^2$, P.~Shagin$^{10}$, Yu.M.~Shatunov$^3$,
E.P.~Sichtermann$^{13}$, E.~Solodov$^3$, M.~Sossong$^8$,
A.~Steinmetz$^{13}$, L.R.~Sulak$^{1}$, C.~Timmermans$^{10}$,
A.~Trofimov$^1$, D.~Urner$^8$, P.~von~Walter$^7$,
D.~Warburton$^2$, D.~Winn$^5$, A.~Yamamoto$^9$ and
D.~Zimmerman$^{10}$ \\
(Muon $(g-2)$ Collaboration)
}

\affiliation{
\mbox{$\,^1$Department of Physics, Boston University, Boston, MA 02215}\\
\mbox{$\,^2$Brookhaven National Laboratory, Upton, NY 11973}\\
\mbox{$\,^3$Budker Institute of Nuclear Physics, 630090 Novosibirsk, Russia}\\
\mbox{$\,^4$LEPP, Cornell University, Ithaca, NY 14853}\\
\mbox{$\,^5$Fairfield University, Fairfield, CT 06430}\\
\mbox{$\,^6$ Kernfysisch Versneller Instituut, University of Groningen,} \\
\mbox{NL-9747 AA, Groningen, The Netherlands}\\
\mbox{$\,^7$ Physikalisches Institut der Universit\"at Heidelberg, 69120 Heidelberg, Germany}\\
\mbox{$\,^8$ Department of Physics, University of Illinois at Urbana-Champaign, Urbana, IL 61801}\\
\mbox{$\,^9$ KEK, High Energy Accelerator Research Organization, Tsukuba, Ibaraki 305-0801, Japan}\\
\mbox{$\,^{10}$Department of Physics, University. of Minnesota.,
Minneapolis, MN 55455}\\
\mbox{$\,^{11}$ Science University of Tokyo, Tokyo, 153-8902, Japan}\\
\mbox{$\,^{12}$ Tokyo Institute of Technology, 2-12-1 Ookayama, Meguro-ku, Tokyo, 152-8551, Japan}\\
\mbox{$\,^{13}$ Department of Physics, Yale University, New Haven,
CT 06520}\\
\mbox{$\,^{\dagger}$ Deceased}\\
}

\date{\today}

\begin{abstract}

Three independent searches for an electric dipole moment
(EDM) of the positive 
and negative muons
have been performed, using spin
precession data from the muon $g-2$ storage ring at 
Brookhaven National Laboratory. Details on the experimental apparatus and 
the three analyses are presented.
Since the individual results on the positive and negative muon, as well
as the combined result,
$d_{\mu}  =(-0.1\pm 0.9) \times 10^{-19}~e\cdot $cm, 
are all consistent with
zero, we set a new muon EDM limit, 
$|d_{\mu}| < 1.9 \times 10^{-19}~e\cdot $cm
($95\%$~C.L.). This represents a factor of 5 improvement over the previous
best limit on the muon EDM.

\end{abstract}

\maketitle


A permanent electric dipole moment (EDM) for a particle in a non-degenerate 
state
violates both parity ($P$) and time reversal ($T$)
symmetries. Assuming conservation of the combined symmetries
$CPT$, where $C$ refers to charge conjugation symmetry, $T$ violation implies
$CP$ violation. Unlike parity violation, which is maximal 
in weak leptonic
decays, CP violation has never been observed in the leptonic sector. 
Given the small
observed levels of $CP$ violation (seen only in the decays of
neutral 
kaons and $B$-mesons), the standard model (SM) 
predicts that the EDMs of the
leptons
are so small that their detection is well beyond
experimental capabilities for the foreseeable future.
Any non-zero experimental value would therefore indicate 
the existence of new physics.
This distinguishes leptons from strongly interacting
particles, which could have a measurable SM EDM if the
$\bar \theta$ parameter in the QCD Lagrangian turned out to be
sufficiently large.
Most models purporting to explain the baryon asymmetry of the universe
require CP violation~\cite{sakharov}, but the 
observed level in the $B$-meson and kaon systems is insufficient, suggesting
that there must be other, as yet undetected sources of CP violation, which
could produce non-vanishing EDMs.
Consequently, experimental searches for EDMs are
being widely pursued. The muon is of special interest because,
again for the foreseeable future, it is the only
particle, outside the first generation, for which a precision 
EDM measurement is 
feasible ~\cite{farley}.

\begin{table*}[t]
\caption{Some experimental limits on EDMs.}
\label{tab:thebest}
\begin{center}
\renewcommand{\arraystretch}{1.2}
\begin{tabular*}{0.6\textwidth}{|p{0.2\textwidth}|p{0.25\textwidth}|p{0.1\textwidth}|}
\cline{1-3}
Physical System &  Value, Error ($e\cdot$~cm) & Reference \\
\cline{1-3}
$^{199}$Hg atom & $(0.49\pm 1.29\pm 0.76) \times 10^{-29}$ & ~\cite{mercurylimit}\\
electron &  $(0.69\pm 0.74) \times 10^{-27}$ & ~\cite{electronlimit}  \\
neutron &  $(-1.0\pm 3.6) \times 10^{-26}$ & ~\cite{neutronlimit} \\
muon & $(-3.7\pm 3.4) \times 10^{-19}$ & ~\cite{BaileyEDM}\\
\cline{1-3}
\end{tabular*}
\end{center}
\end{table*}

\par

The best experimental limits on the EDMs of elementary particles or atoms,
some of which are listed in Table~\ref{tab:thebest},
are those for the $^{199}$Hg atom, the
electron, and the neutron. The measured values
are all consistent with zero. These stringent limits, which are likely
to improve over the next few years,
have produced some of the most significant constraints
on extensions to the SM, many of which predict relatively large
EDMs ~\cite{thetabartheory1,thetabartheory2,thetabartheory3,thetabartheory4}.
The data on the neutron lead to the current 
limit ~\cite{thetabarexp}, 
$\bar \theta < 10^{-10}$.
\par
The current limit on the muon EDM ~\cite{BaileyEDM}, set by the last muon
$g-2$ experiment at CERN and also shown in Table~\ref{tab:thebest}, 
is considerably less stringent.
In the SM and in some of its extensions, the lepton EDMs scale
with mass. Scaling the measured electron EDM by the ratio of the muon to the
electron
mass implies a limit for the muon of 
$d_{\mu}=(1.4\pm 1.5) \times 10^{-25}$ $e\cdot$cm.
However, the scaling which could come from new physics is essentially
unknown. Indeed, some SM extensions predict that the muon EDM is
larger than $10^{-23}$ $e\cdot$cm~\cite{babu}. 

While the primary objective of the BNL Muon $(g-2)$ experiment 
~\cite{carey,brown1,brown2,bennett1,bennett2,PRD}
was to measure the anomalous magnetic dipole moment of the muon, $a_{\mu}$,
this paper describes how, as a secondary measurement, a new limit on the 
electric dipole moment of the muon
is obtained, a factor of 5 improvement over that achieved by the
CERN experiment.
The improved limit is interesting in its own right and
helps to resolve  one ambiguity in interpreting the difference between the 
theoretical and measured values of $a_{\mu}$.
\par
In general, the electric dipole moment, $\vec d$, and magnetic dipole 
moment~(MDM, $\vec \mu$) are given by
\begin{xalignat}{2}
\vec d = \eta {\frac{q \hbar}{4mc}} \vec S & \qquad \vec \mu = g{\frac{q}{2m}} \vec s,
\label{eq:edmmdm}
\end{xalignat}
\noindent
where $\eta$ describes the size of the EDM, $gq/(2m)$ is the 
gyromagnetic ratio, 
$q$ and $m$ are the particle's charge and mass, respectively, and ${\vec S}$ 
is a unit vector directed along $\vec s$, the true spin vector. 
For muons circulating in a storage ring, such as those used by both the CERN
and BNL experiments, 
the spin vector's precession in the muon rest frame (MRF)
depends on the interaction of its MDM 
with the magnetic field and of its EDM with the
electric field.
Measurements involving the former interaction can provide a
determination of the magnetic anomaly,
while those involving the latter can be used  to determine the EDM.
Note that if CPT symmetry is assumed, the $\eta$ parameter
is the same for positive and negative muons while ${\vec d}$ changes sign.


\par
In the presence of electric and magnetic fields, in the laboratory frame
of reference,
the rate of spin precession relative to the muon momentum direction 
is given by the
sum of contributions from the MDM and EDM:
${\vec \omega}=\vec \omega_a+\vec \omega_{{\text {EDM}}}$,
where, in the approximation $\vec \beta\cdot \vec B\approx 0$,

\begin{equation}
\vec \omega_a=-{\frac {q}{m}}
\biggl[a_{\mu}\vec B
+ \biggl( -a_{\mu}+{\frac{1}{\gamma^2-1}}\biggr)
\vec \beta \times {\frac{\vec E}{c}}\biggr].
\label{eq:omegaa}
\end{equation}

\noindent
In the approximation $\vec \beta\cdot \vec E\approx 0$,

\begin{equation}
\vec \omega_{{\text {EDM}}}=-\eta{{\frac {q} {2m}}}{\biggl(\vec \beta \times \vec B
+{\frac{{\vec E}} {c}}\biggr)}=-{\frac{\eta}{2mc}}\vec F,
\label{eq:omegaedm}
\end{equation}
\noindent
where $q$ is the muon charge, $\vec B$ is the main dipole field and 
$\vec E$ is the field of the
focusing electrostatic quadrupoles,
$a_{\mu}={(g-2)/ 2}\approx 10^{-3}$ is 
the anomalous
magnetic moment and $\vec F$ is the Lorentz force. 
Here, $\omega_a$, the $g-2$ precession frequency,
includes contributions from both the Larmor and Thomas precessions. 
Both $\vec F$ and therefore
$\vec \omega_{\text{EDM}}$
are oriented along the radial direction while the magnetic field and therefore
$\vec\omega_a$ are directed vertically. A non-zero EDM slightly tips 
the direction
of $\vec \omega_a$ out of the vertical direction and slightly increases the 
precession frequency.
\par
Under a Lorentz 
transformation, the laboratory
magnetic field leads 
to a large electric as well as a magnetic field in the MRF.
In fact, since
the effect of the induced electric field, from $\vec \beta \times \vec B$,
on $\vec \omega_{EDM}$
is much larger than that produced by 
the focusing quadrupole field, the latter is ignored.

The BNL $(g-2)$ experiment 
and the earlier CERN\cite{cern3a,cern3b}
experiment were optimized to measure $a_{\mu}$.
Both experiments stored muons with the ``magic'' momentum $p\approx 3.094$
GeV/c, 
corresponding to  $\gamma\approx 29.3$ and $\gamma \tau \approx 64.4~\mu$s,
which gives
$ (\gamma^2-1)^{-1}-a_{\mu}\approx 0$. With this choice of beam momentum, the
effect of the focusing electric field on ${\vec \omega_a}$ is zero. 
$\vec \omega_a$ is oriented vertically, parallel or anti-parallel to $\vec B$.

%

\par
Both the MDM and EDM measurements rely on the detection of positrons or
electrons from the three-body decays of the muons, 
$\mu^+\rightarrow e^++\nu_e+\bar \nu_{\mu}$ or 
$\mu^-\rightarrow e^-+\bar \nu_e+ \nu_{\mu}$.
Positive muons were stored during the 1999 and 2000 data runs, while negative
muons were stored in the 2001 data run.
(Positrons will be used generically to refer to both electrons and positrons in
the subsequent discussion.) The positron laboratory energies range from 0 to
$3.1$~GeV.  All positrons with energies in excess of 1.2 GeV, 
which are the ones of
interest here, are initially 
directed within approximately 40 mrad of the laboratory muon momentum 
direction.  Most
positrons have momenta that are less than those of the muons, 
and are swept by the
magnetic field to the inside of the storage ring, where they can be intercepted
by the detector system.

A consequence of maximal parity violation in the muon decay
is a 
large correlation between the
direction of the positron momentum and the muon spin in the MRF.
For $\mu^{+}$
($\mu^{-}$) decay, the positron (electron) momentum is preferentially 
directed parallel
(anti-parallel) to the spin.

Transforming to the laboratory frame, the number of positrons in a given energy
range is modulated
according to whether the polarization is along or opposite the
direction of motion of the muon, following the familiar functional form

\begin{equation}
N(t)=N_0e^{-\lambda t}(1+A\cos{(\omega t +\Phi)}),
\label{eq:par5}
\end{equation}

\noindent where
$\tau_{\mu}=\gamma \tau_0={\lambda}^{-1}$ is
the dilated muon lifetime. 
For the $\mu^+$ ($\mu^-$), $N(t)$ is a maximum
when the muon polarization is parallel (anti-parallel) to the muon
momentum.
The observed value of $\omega$ is used to deduce
$a_{\mu}$, under the assumption that
the effect of the EDM on the magnitude
of the spin
precession can be neglected. See the discussion associated with 
Eq.~(\ref{eq:biggerwa}) below.

\par
For a uniform B field and in the absence of an EDM,
all muon spins would precess at the same rate, 
regardless of trajectory or momentum, except for very small corrections
for betatron motion and for the effect of the electric field on those
muons whose momenta are not precisely ``magic''. The precession vector
$\vec\omega$ would be anti-parallel to the vertical magnetic field
or, equivalently, the spin vectors would 
precess in the horizontal plane. To investigate the implications of a 
non-zero EDM, the detector acceptance and 
spin motion are described in more detail.

The five-parameter equation for $N(t)$,
Eq.(\ref{eq:par5}), can be rewritten in the more general (differential) form,
using MRF coordinates 
\begin{equation}
P(\alpha,{\tilde{\theta}},{\tilde{\phi}},t) d\alpha d{\tilde{\Omega}} = n(\alpha)\biggl[1+a(\alpha){\hat{\tilde{p}}}\cdot{\hat{s}}(t)\biggr]d\alpha d{\tilde{\Omega}},
\label{eq:par5moddiff}
\end{equation}
where $\alpha = (E/E_{max})^{\text{MRF}}$, the fractional positron decay energy
in the MRF, 
${\hat{\tilde{p}}}$ is a unit vector along the decay positron's momentum, 
$\hat s$ is the muon spin and $a$ is the decay asymmetry. The tildes refer
to MRF coordinates: ${\hat{\tilde{x}}}$ points in the
outward radial direction, ${\hat{\tilde{y}}}$ points upward, and 
${\hat{\tilde{z}}}$ lies along the muon momentum. The corresponding
polar and azimuthal angles are denoted ${\tilde\theta}$ and ${\tilde\phi}$,
respectively.
The decay asymmetry is given by the expression
\begin{equation}
a(\alpha) = { \frac {2 \alpha -1}{3 - 2\alpha}}.
\end{equation}
Alternatively, in integral
form, we have
\begin{equation}
N(t)=N_0e^{-\lambda t}(1+\vec A\cdot \vec s(t)).
\label{eq:par5mod}
\end{equation}

\noindent
where $\vec s$ is the precessing polarization vector of the muon
ensemble, $\vec A$ is the average {\it asymmetry vector},
the average of $a{\hat {\tilde p}}$ over detected decays.
Of course, the asymmetry vector of a data sample depends on the 
MRF energies and angles of {\it accepted} positrons.
For example, if positrons were only accepted if they struck
detectors above~(below) the storage ring midplane, 
then $\vec A$ would acquire a
positive (negative) vertical component.
\par
In Eq. ~(\ref{eq:par5mod}), ${\vec s(t)}$ describes the spin motion. ${\vec A}$
includes the effects 
of the weak decay on the electron distribution, as 
well as detector acceptance. 
Consider the $a_{\mu}$ measurement. We define a local, lab-frame, 
Cartesian coordinate system where $\hat{x}$ and $\hat{y}$ are 
directed radially
outward and vertically up, respectively and $\hat z$ is 
longitudinal, along the motion of the stored muon beam.
For the $a_{\mu}$ analysis, there is no bias in the spectra of positrons
selected,  
above or below the midplane, and therefore
$A_y\approx 0$. Moreover, there is little detector bias  for
positrons
with negative versus positive radial momentum components. Thus 
$A_x \approx 0$. 
The preferential selection of positrons with high laboratory frame energies 
results in a large value of $A_z$. 
Therefore
$\vec A\approx A_z \hat z$, with $\hat z$ longitudinal and 
Eq.~(\ref{eq:par5mod})
reduces to Eq.~(\ref{eq:par5}). 
\par
\begin{figure}[h]
\includegraphics[width=0.48\textwidth]{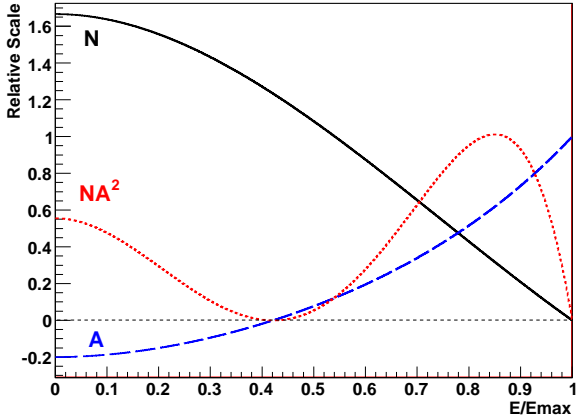}
\caption{Number $N$ (solid), $g-2$ differential asymmetry $A$(dashed) and 
$g-2$ statistical figure of merit (FOM, dotted) $NA^2$ vs. laboratory frame energy fraction.}
\label{fig:g2fom}
\end{figure}

For a lab-frame decay electron energy 
fraction $f = E/E_{max}$, $A_z$ is given by
\begin{equation}
A_z = {\frac {8f^2-f-1}{5+5f-4f^2}},
\end{equation}
with a corresponding (relative) number of muons 
\begin{equation}
N = {\frac {2(1-f)(5+5f-4f^2)}{12\pi}},
\end{equation}
ignoring the effect of detector acceptance.
The magnetic anomaly $a_{\mu}$ is derived from the experimentally determined
value of $\omega$. 
The statistical uncertainty on $\omega$ in Eq.~(\ref{eq:par5}) 
is inversely proportional to $(N_0A_0^2)^{1/2}$ where $N_0$ and $A_0$ refer
to the ensemble of accepted events.
Graphs of differential $A$  (or $A_z$, strictly speaking), $N$  
and $NA^2$ vs. $f$, the last a 
statistical figure
of merit (FOM), are shown in Fig.~\ref{fig:g2fom}. Imposing an energy threshold
selects a subset of decay positrons with a net average longitudinal momentum
in the MRF, which leads to the oscillation in $N(t)$.
%

\par
\begin{figure}[h]
\includegraphics[width=0.48\textwidth]{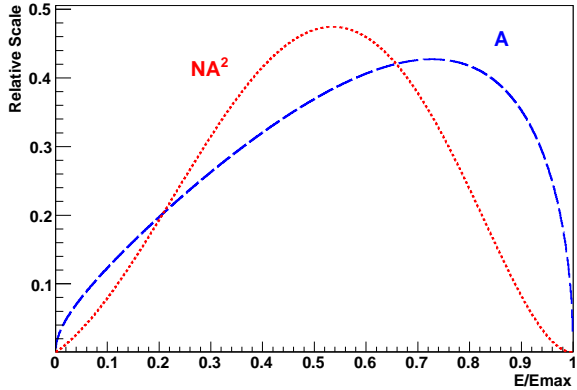}
\caption{Differential EDM asymmetry, $A_u$ (dashed), and statistical figure of merit, $N A_u^2$ (dotted), vs. laboratory frame energy fraction.}
\label{fig:edmfom}
\end{figure}

The asymmetry vector formalism can also be used in the context of the EDM
analysis.  
The asymmetry 
for decays selected along any direction $u$ in the transverse ($x,y$) plane 
is given by
\begin{equation}
A_u = \cos \phi {\frac {8(1+4f){\sqrt{(1-f)f}}} {5(5+5f-4f^2)}},
\end{equation}
where $\tan \phi =  p_y/p_x$, with $p_x$ and $p_y$ the transverse momenta
and $\phi$ the azimuthal angle.
The corresponding asymmetry, $A_u$, and differential statistical figure of 
merit, $N A_u^2$, are shown in Fig.~\ref{fig:edmfom}. The sensitivity to an
EDM is greatest over a broad range of lab energies, for  $0.2$ to ~$0.7$ of
the maximum.

\par
Concerning the spin motion, with no EDM, the only significant 
torque is that produced by 
the interaction of the magnetic moment with
the magnetic field.
The spin vector precesses in the x-z plane
relative to the momentum vector according to
$\vec \omega_a=-a_{\mu}(q/m)\vec B$.
For $\mu^+$, the polarization vector as a function of time is given by

\begin{equation}
\vec s(t)=-s_{\perp} \sin{(\omega t+\Phi)}\hat {x}
-s_{\parallel}\hat y +s_{\perp}\cos{(\omega t+\Phi)}\hat z,
\end{equation}

\noindent
where $s_{\parallel}$  and 
$s_{\perp}$ are, respectively, the magnitudes of the components 
parallel and perpendicular
to $\vec \omega$.


\par
With a non-zero EDM, the precession is no longer confined to
the horizontal plane.
When the torque due to the motional electric field acting on an
EDM is included, there is an additional spin precession
$\vec \omega_{\text{EDM}}=-(\eta /2) (q/m) \vec \beta\times \vec B$, which
is directed radially in the storage ring.
The plane of spin precession, which is perpendicular to the total
precession vector $\vec \omega=\vec \omega_a+\vec \omega_{\text{EDM}}$,
is tipped out of the orbit plane by an angle
\begin{equation}
{\delta}=\tan^{-1}[\eta \beta /(2a)], 
\label{eq:tipping}
\end{equation}
where $a$ is the anomaly and we see that 
$\delta$ is approximately proportional to the magnitude
of the EDM. 
The spin precession is now given by
\begin{eqnarray}
\vec s(t) & = &
(-s_{\perp}\cos{\delta}\sin{(\omega  t+\Phi)}+s_{\parallel}\sin{\delta})\hat x \notag \\ 
& & - (s_{\parallel}\cos{\delta}+s_{\perp}\sin{\delta}\sin{(\omega t+\Phi)})\hat y \notag \\
& & + s_{\perp}\cos{(\omega t+\Phi)}\hat z,
\label{eq:tipping2}
\end{eqnarray}

\noindent
that is, the average vertical component of the spin polarization
oscillates at angular frequency $\omega$, with an amplitude which is 
proportional to the EDM. The parallel and perpendicular
components are defined with respect to the new direction of ${\vec \omega}$.
If $s_{y0}$ represents the
vertical component of the beam's polarization at injection, 
which in the $g-2$ experiment is very small,
the maximum  excursion $|s_y-s_{y0}|$ occurs when the polarization is 
directed either
along or opposite to the radial direction. To be precise, the vertical EDM
oscillation leads the $(g-2)$ oscillation in $N(t)$ 
by 90~$^\circ$.
This phase
difference, which is the same for positive and negative muons, is useful
in suppressing false EDM signals, as
discussed below. It is also important to note that the phenomenology of the
EDM-related spin motion is charge independent - if $\eta$ is the same for 
positive and negative muons, the tipping of the precession plane and the
small change in frequency are identical. 
\par
If the asymmetry term in Eq. ~\ref{eq:par5moddiff} is given the explicit
form 
\begin{multline}
a(f,\theta)\hat p(f,\theta,\phi) = \\
A_{xy}(f,\theta)\cos\phi{~\hat x} +
A_{xy}(f,\theta)\sin\phi{~\hat y} + A_z(f,\theta){~\hat z},\\
\label{eq:asymmetrypaul}
\end{multline}
using lab-frame variables $f$ and $\theta$, the corresponding dot-product 
takes the form

\begin{multline}
a(f,\theta)\hat p(f,\theta,\phi)\cdot{\hat s}(t) =  \\
A_z(f,\theta)(s_{\perp}\cos(\omega t + \Phi)) +  \\
A_{xy}(f,\theta)\cos\phi (-s_{\perp}\cos\delta\sin(\omega t + \Phi) +
 s_{\parallel}\sin \delta) + \\
A_{xy}(f,\theta)\sin\phi (-s_{\parallel}\cos\delta- s_{\perp}\sin{\delta}\sin(\omega t + \Phi)). \\
\label{eq:dotproduct}
\end{multline}
The first term on the right-hand side, 
which arises from the longitudinal component of the asymmetry vector, has the 
$\cos(\omega t + \Phi)$ time dependence  characteristic of 
the anomalous precession. The second (radial asymmetry) term, 
receives a  contribution from the EDM, but is not readily
observable because it cannot be distinguished from a small shift in 
the oscillation phase angle. 
The third (vertical asymmetry) term has the sought-for vertical oscillation,
with an amplitude proportional to the EDM.
\par
As noted above, a non-zero EDM increases the total spin precession frequency 
\begin{equation}
\omega=\omega_a\sqrt{1+\tan^2{\delta}}.
\label{eq:biggerwa}
\end{equation}
\noindent
While the roughly 3 standard deviation difference between the measured and 
theoretically-predicted values of $a_{\mu}$ ~\cite{PRD} could be the result 
of new physics such as supersymmetry or muon substructure, it could also 
be caused by a shift in the value of $\omega_a$ produced by a non-zero EDM.
See Eqs. ~(\ref{eq:omegaa}), ~(\ref{eq:omegaedm}) and ~(\ref{eq:biggerwa}).
A sufficiently precise measurement of the muon EDM could help
resolve this ambiguity.
\section{Experimental Apparatus}
\begin{figure}[h]
\includegraphics[width=0.48\textwidth]{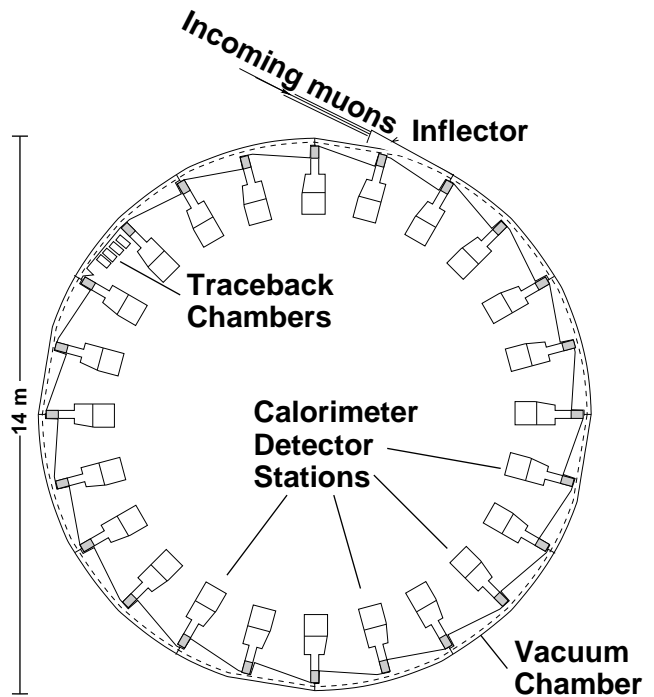}
\caption{A plan view of the $g$-2 storage ring showing the positions of the 24
calorimeters, the traceback chambers and other devices. The calorimeters
are numbered 1-24 starting from the injection point and proceeding clockwise.}
\label{fig:g2layout}
\end{figure}

Details on the $g-2$ storage ring ~\cite{danby}, 
detectors ~\cite{sedykh,prisca} 
and on the anomalous
precession analysis are presented
elsewhere~\cite{carey,brown1,brown2,bennett1,bennett2,PRD}.
The storage ring
magnet is 44 m in circumference with a dipole field of 1.45 T.
 A plan view of the storage ring and detectors is shown in 
Fig.~\ref{fig:g2layout}.
Muons are injected nearly tangent to the circumference
through the field-free region 
provided by a superconducting inflector.
The beam is centered on the storage region
by means of a small magnetic kick applied at 90$^\circ$ downstream 
from the inflector.
Data used in the anomalous precession and
EDM analyses are collected for more than
ten muon lifetimes during each injection
cycle. While the details of beam injection and magnet design were
rather different, the layout of the CERN III storage ring and 
detectors were very similar to those of E821.

\section{Measuring the EDM}
The measurement of the muon EDM requires the detection of
oscillations in $p_y$, the vertical momentum of the decay positrons,
which reflect the oscillations in $s_y$, the vertical polarization of
the muon beam. The amplitude of the oscillations is
proportional to the EDM. A measurement of the average vertical decay angle
vs. time, which picks out the third term on the right-hand side of  
Eq. (\ref{eq:dotproduct}), 
provides the most direct indication of possible oscillations
in $s_y$.  

In order to optimize the experimental sensitivity to a non-zero EDM,
the range of positron energies should maximize
the FOM for oscillations in $s_y$ (see Fig ~\ref{fig:edmfom}) and
maximize $s_{\perp}$. Naturally,
the beam polarization is optimized for the
$g-2$ measurement. The EDM FOM varies slowly with positron energy - 
a broad range of energies around 1.5 GeV is acceptable. It is important 
to note, however, that the acceptances
of the $(g-2)$ detector systems are not optimized for the EDM measurement.
Those acceptances generally rise with increasing positron energy,
reach a maximum near 2.8 GeV and 
then fall to the energy endpoint at 3.1 GeV.
\par 
If direct tracking measurements are unavailable (as was the case in the
CERN III experiment), the observation of certain rate oscillations with
a $\sin(\omega t + \Phi)$ time dependence (see Eq. (~\ref{eq:dotproduct}))
can be used instead. In
this case, one should
select data subsets with non-zero average positron momentum along 
$y$, in order to maximize $|A_y|$.
In practice, this means selecting decay positron hits either above or
below the storage ring midplane. A further
refinement is to measure the 
oscillation phase as a function of 
vertical position on a detector. The $\sin(\omega t + \Phi)$ time
dependence of the vertical oscillation
is shifted by 90 degrees
relative to the $g-2$ precession ($\cos(\omega t + \Phi)$) and the $\sin \phi$
term provides a sign flip between the signals observed above or below
the storage ring midplane.
In both these approaches, the correlation of detected vertical position 
to vertical angle, while
strong, is reduced somewhat by the range of vertical decay positions. 
\par
The CERN III
experiment \cite{BaileyEDM} mounted two
adjacent scintillator paddles on the entrance face (where most
positrons enter) of one of
the calorimeters.
One paddle was mounted above the vertical mid-plane of the
calorimeter, and the other below.  An event was counted when a signal from a
paddle was registered 
in coincidence with a calorimeter signal that exceeded a specified
energy threshold, typically $1.2-1.4$ GeV.  Any oscillation in 
$\langle p_y \rangle$ would
cause a corresponding oscillation in the ratio of the sum and difference of 
count rates in the up $(N^+)$ and down $(N^-)$
paddles,

\begin{equation}
r(t)={\frac {N^{+}(t)-N^{-}(t)}{ N^{+}(t)+N^{-}(t)}}.
\label{ratio}
\end{equation}
In the presence of an EDM but in the absence of field perturbations which
could  bend decay electron tracks,
expressions for $N^{+}$ and $N^{-}$ can be written in terms of
the asymmetry vector for each data sample.
A sub-sample of decay positrons which are 
detected above the orbit plane will have a
non-zero vertical asymmetry component, $\vec A^+=A_y \hat y+A_z\hat
z$. 
Similarly a sub-sample below the orbit plane will have $\vec A^-=-A_y \hat
y+A_z\hat z$.
Assuming the gain and acceptance of the upper
and lower detectors are equal and the storage ring and vertical detector
midplane are identical, Eqs.~(\ref{eq:par5mod}) and ~(\ref{eq:tipping2}) 
with $s_{\parallel} = 0$ give
\begin{eqnarray}
\lefteqn{N^{\pm} = {\frac{1}{2}}N_0e^{-\lambda t} \times}   \notag \\
& & (1\mp A_y s_{\perp}\sin{\delta}\sin{(\omega t+\Phi)}+A_zs_{\perp}\cos{(\omega t+\Phi)}). \notag \\ 
\end{eqnarray}


\noindent
As expected, the term containing $A_y$ is proportional to the EDM, 
oscillates at
angular frequency $\omega$ and is $90^\circ$ out of phase with the ordinary
$g-2$ precession given by the term containing $A_z$.

Separating out the contribution from the EDM, the
expression for $N^{\pm}$ becomes
\begin{equation}
N^{\pm}={\frac {1}{2}}N_0e^{-\lambda t}
[1\mp A_{{\text {EDM}}}\sin{(\omega t+\Phi)} + A_{\mu}\cos{(\omega t+\Phi)}],
\end{equation}

\noindent
where $A_{{\text {EDM}}}=A_ys_{\perp}\sin{\delta}$ is proportional to $d_{\mu}$ and
$A_{\mu} = A_z s_{\perp}$.

Equivalently one can define an angle $\Psi=\tan^{-1}{(-A_{{\text {EDM}}}/A_{\mu})}$
and an overall asymmetry $A=\sqrt{A_{\mu}^2+A_{\text {EDM}}^2}$
and write
\begin{equation}
N^{\pm}={\frac{1}{2}}N_0e^{-\lambda t}[1+A\cos{(\omega t+\Phi\pm\Psi)}].
\label{eq:par5pm}
\end{equation}

\noindent
In the presence of an EDM only (again, without field perturbations), 
$d_{\mu}$ can be obtained
from either a fit to the ratio, Eq.~(\ref{ratio}),

\begin{equation}
r(t)={\frac{{A_{{\text {EDM}}}\sin{(\omega t +\Phi)}}}   {1+A_{\mu}\cos{(\omega t +\Phi)}}},
\label{eq:ratio1}
\end{equation}

\noindent
or from separate fits of Eq.~(\ref{eq:par5pm})
to the data from the top and bottom paddles, with the EDM being inferred from
the magnitude of angle $\Psi$. The latter approach is better, since many
small spin perturbations, for example from spin resonances, change
$N^+-N^-$ in Eq. ~(\ref{eq:ratio1}) without changing $\Psi$ in 
Eq. ~(\ref{eq:par5pm}). Spin resonances, however, are very weak in the
$g-2$ storage ring. Only high longitudinal modes of some
non-linear field components can oscillate, in the MRF, in resonance
with spin precession. For such resonances, the original constant part of
the vertical spin component changes slowly.

Use of the latter approach  led to the CERN
result\cite{BaileyEDM} on the combined $\mu^+$ and $\mu^-$ EDMs,
$d_{\mu}=(3.7\pm 3.4) \times 10^{-19}$ $e\cdot$cm. This is consistent with zero,
giving $d_{\mu}<1.05\times 10^{-18}$ $e\cdot$cm at the 95\% confidence level.
The overall uncertainty is evenly split between statistical and systematic
errors.


\par
The systematic error arises mainly from the uncertainty in the alignment of the
detectors relative to the muon beam, which, coupled with the 
spin precession, can produce a false EDM signal.
Along the trajectory from the muon decay point to the detector,
positrons emitted with outward radial momentum components
will travel
further than those with inward components, and will therefore have more time to
spread out in the vertical direction.  When the muon spin is
pointing radially outward, more positrons are emitted with outward radial
momentum components than inward, and the width of the vertical distribution of
positrons at the detector entrance face will be larger than when the spin is
pointing radially inward. The width of the vertical distribution of positrons
therefore 'breathes' at the spin precession frequency ${\omega / (2\pi)}$.

\par
Ideally, the average distribution of the muons in the 
vertical coordinate is symmetric 
about $y=0$, 
and this symmetry is generally reflected in
the decay positrons observed by the detectors. Indeed, if the EDM is zero, if
the initial vertical component of the spin polarization is zero,
if the vertical position of the dividing line between scintillator paddles
is aligned with $y=0$, and if
the paddles have the same
efficiency of positron detection, then the vertical distribution of detected
positrons is symmetric in $y$ at all times and $r(t)=0$.
If any of these requirements are not met, then the time average value
of $r(t)$, $\langle r(t) \rangle$,
will be non-zero in general.
A non-zero $\langle r(t) \rangle$ is not, by itself, significant. However, 
the breathing of the vertical width of the beam will now introduce oscillations in $r(t)$ 
at frequency ${\omega/ (2\pi)}$, {\it in phase with the oscillations that
would be produced by a true EDM}. In other words, it would produce a false EDM signal.



\par
Corrections for detector inefficiency and misalignment can be made,
using knowledge 
of the shape of the vertical distribution as a function of time,
the acceptance of
the paddles as a function of time, the value of $s_{z0}$, and the
measured value
of $\langle r(t) \rangle$. However, in the case of the CERN III experiment, which had
only two paddles mounted on a single calorimeter,  detailed measurements 
of the vertical distribution 
could not be made.
In E821, there is more information on the vertical distribution because
the detectors have more than two-fold vertical segmentation. 
However, the relative acceptances
of the elements are more 
complicated because they depend not only on the efficiency of the scintillators
but also on the gain stability and relative efficiency, versus time, 
of the top half of the
calorimeter
relative to the bottom.  The time dependence of the vertical distribution on
the detector face is readily simulated, but not the detector response, which
must take into account the time-varying energy and incident angles of the
positrons, as well as geometric variations in gain and efficiency of the 
scintillator and calorimeter.  Limitations on the corrections cause these
alignment effects to dominate the systematic error.


%

\section{The BNL EDM Measurements}

\begin{figure}
\includegraphics[angle=90,width=.48\textwidth]{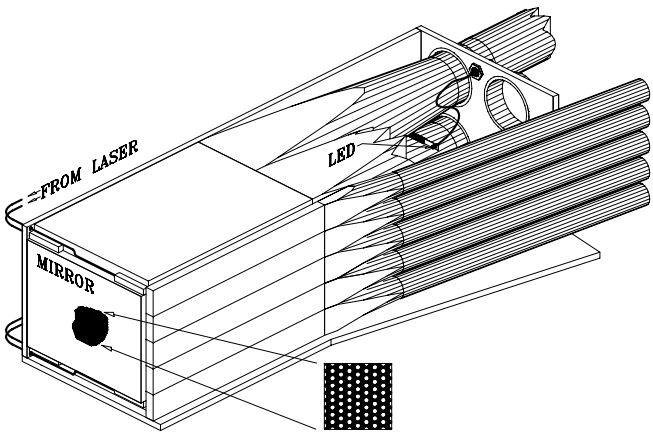}
\caption{Detail view of a detector station. The calorimeter consists of
scintillating fiber embedded in a lead-epoxy matrix, with the fibers being
directed radially in the storage ring. Five horizontal scintillators, the 
FSD segments, cover the positron entrance face of the
calorimeter. Each calorimeter is approximately 23~cm (wide) by 
14~cm (high) by 16~cm (deep). 
The PSD (not shown) is placed in front of the FSD.}
\label{fig:caloblowup}
\end{figure}

\begin{figure*} 
\includegraphics[width=\textwidth,angle=0]{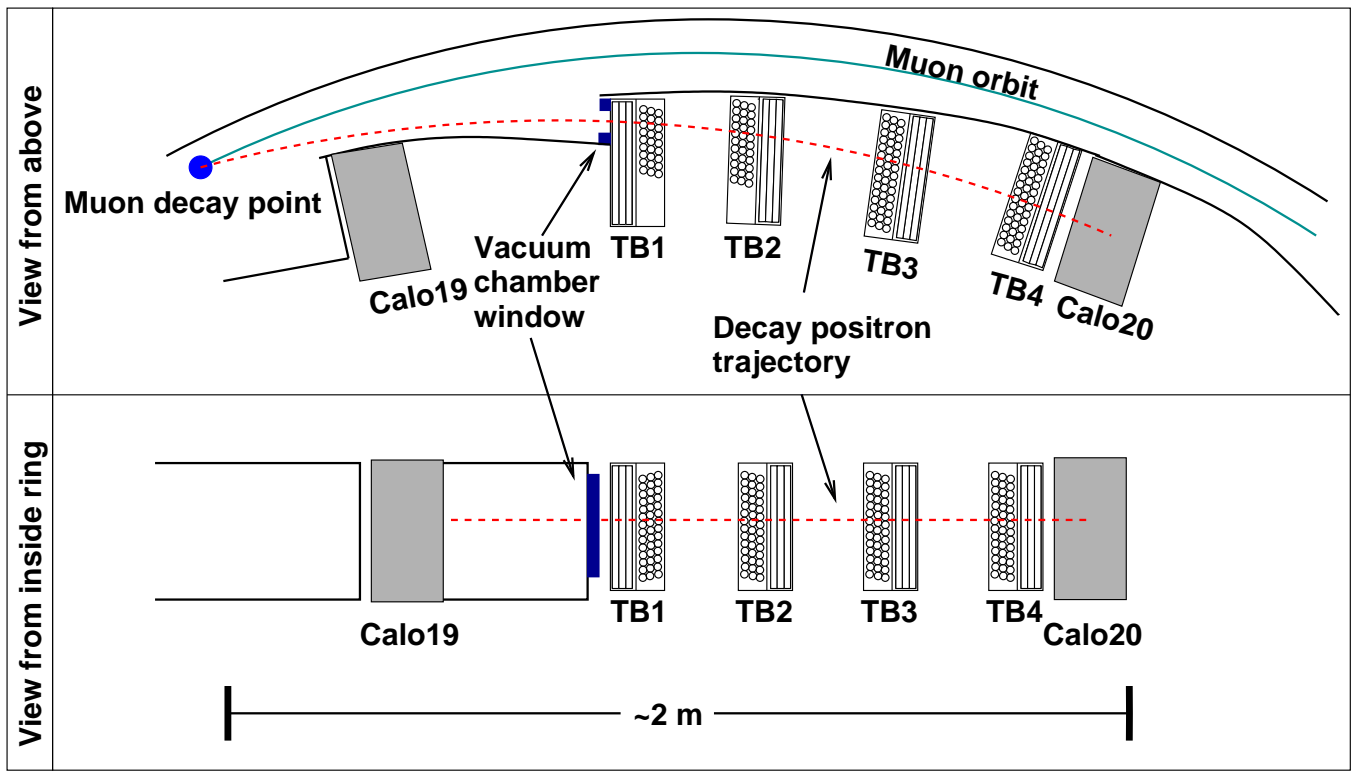} \hfill
\caption{Top and side views (not to scale) of the traceback system.
Muons decay
to positrons in the storage region.  The positron then must travel through the
thin window in the vacuum chamber scallop, through the traceback chambers
and into the
calorimeter.  Horizontally lying straw chambers
allow the precise reconstruction the
vertical angle of the positron.}
\label{fig:tbackdrawing}
\end{figure*}

In addition to the positron calorimeters, three kinds of detectors were used 
to measure the EDM in the BNL experiment:

\begin{itemize}

        \item The front scintillation detectors (FSD) are used to measure any
        oscillation in the average vertical position of the decay positrons
        as they enter the calorimeters.  
        They consist of an array of five horizontal
        scintillator paddles which cover the entrance faces of 
	roughly half the calorimeters ~\cite{sedykh}. 
	See Figs. ~\ref{fig:g2layout} and ~\ref{fig:caloblowup}. 
        About 10 of the FSDs were instrumented, typically those 
        mounted on calorimeters where the injection-related
        background was the least severe.
        The increased segmentation over the CERN arrangement
        helps improve
        knowledge of the vertical distribution of the positrons,
        thereby reducing the
        misalignment error, and permits the more sophisticated analysis
        which is described below.

        \item The position sensitive detectors ~\cite{prisca} (PSDs) 
        are a much more
        finely segmented
        version of the FSD, with horizontal as well as vertical segmentation.
        They
        cover the  positron entrance faces of calorimeters 13, 14, 15, 18 and
        24.
        \item The traceback wire chamber system (TWC) consists of a series
        of 
        drift chambers mounted in front of calorimeter 20, along the
        path of the 
        incoming decay positrons. See Fig. ~\ref{fig:tbackdrawing}. 
        The TWCs are used to measure the pitch angle of the decay positrons.
        They also provide information on the phase
        space distribution of the muons in the storage ring.
        Tracks measured by the traceback system
        are extrapolated back into the storage volume, to the position where
        the momentum
        points tangent to the storage ring, which is a good approximation to
        the point of decay.
\end{itemize}

%
 

\par

%
\par
Another difference between the CERN III and BNL EDM measurements concerns
the stored beam. The functions which describe the time spectra (Eq. 
(~\ref{eq:par5mod}), for example) are constructed with the assumption 
that while the 
muons themselves are moving, the spatial distribution is static.
In fact, the muon beam arrives in a bunch, which provides a modulation of
the decay signal with period 149.2 ~ns, the time it takes for the bunch
to circle through the storage ring. This
{\it fast rotation} signal, which is prominent for roughly the first 
70 ~$\mu$s,
was filtered out in both the CERN III and E821 
analyses. However, because of our direct muon injection technique, there is
another very important collective beam motion. For several hundreds of 
microseconds after injection, the stored 
muon beam exhibits a variety of coherent betatron oscillations collectively 
referred to as CBO. Among these is an oscillation in the radial beam
position (measured at a fixed point in the ring) at frequency
$f_{cbo}\approx 465$~kHz~\cite{PRD}.  
Descriptions of the vertical distributions of positrons on detector faces 
must account for beam motion.
When the beam radius is larger than average, the width
of the vertical distribution of positrons at the detector is larger because of
the longer distance traveled, and conversely, when the beam radius is smaller
than average, the vertical width is smaller.  
\par
Combined with a detector-beam
misalignment, radial beam motion can produce the same sort of vertical
oscillation in $\langle p_y \rangle$ as the anomalous precession of the spin, 
again proportional to the
magnitude of the misalignment.
However, because the CBO-induced frequency is different from that produced 
by an EDM,
this oscillation will not be
mistaken for an EDM signal, but instead can be used to calibrate,
and ultimately to correct, the misalignment error.
\par
While the FSD and PSD provide EDM measurements patterned on that of the
CERN III
experiment, the PSD provides another, qualitatively
different, approach. 
The phase parameter $\Phi$ in Eq.~(\ref{eq:par5}), 
varies with $y$, the detector vertical
position. As explained earlier, for decays in which the polarization points
outward, the ensemble of decay
positrons spread out more in the vertical direction
than when the polarization points inward.  As a consequence, positrons detected
far away from the vertical mid-plane ($|y|$ large) will be more likely to come
from outward decays than from inward decays.  Conversely, inward decays will be
more likely than outward decays when $|y|$ is small.  Therefore $\Phi$ will
depend slightly on $y$. If $d_{\mu}=0$, the plane of the spin precession is in
the horizontal plane, and $\Phi(y)$ will be symmetric in $y$.
However, if $d_{\mu}\neq 0$,
the plane of the spin precession is tipped out of the horizontal plane, and
$\Phi(y)$ will {\it not} be symmetric in $y$.  
\par
Finally, although the TWCs were originally designed to determine
the phase space of the stored muons, for use in the anomalous precession
analysis, their measurement of the average vertical decay angle provides
an independent measurement of $d_{\mu}$, one which is largely immune to the
detector misalignment problem.
In the absence of radial magnetic fields, the vertical angle of the track 
as measured in the traceback system, is the same as that at the moment of
decay. A non-zero EDM would be reflected in an oscillation of the vertical 
component of the positron momentum, $90^\circ$ out of phase with the (g-2)
number oscillation.
Further details on the EDM measurements made with these three
detectors are presented in the following sections.

\section{Traceback Analysis of 1999 and 2000 $\mu^+$ Data Sets}
\label{sec:mikesanalysis}

\subsection{The Traceback Detector}

The traceback detector consists of a
set of eight, three-layer drift tube planes,
designed to measure positron trajectories along their usual decay path
out from the storage volume into the calorimeters.
This detector was installed during the 1999 and 2000 running periods at
a single
position in the ring.  By analyzing the positron drift time
spectrum in a
straw, the radius from the anode
wire at which the particle passed is determined. 
Tracks are fit to these drift circles
according to the equations of motion of a charged particle in a magnetic
field. 
Due to the very inhomogeneous magnetic field in the region of the traceback
detector~(10~T/m), tracks must be integrated with a very small step size.  
The relatively high precision of the track reconstruction offers a detailed, 
time
varying picture of the decay positron trajectories and the stored muon beam.
In particular, for the positron's vertical angle of entrance into the detector,
which is examined for an EDM oscillation signal,
the resolution is approximately 350~$\mu$rad.

\subsection{Traceback analysis description and results}

\begin{figure}[h] 
\begin{center}
\includegraphics[width=0.49\textwidth,angle=0]{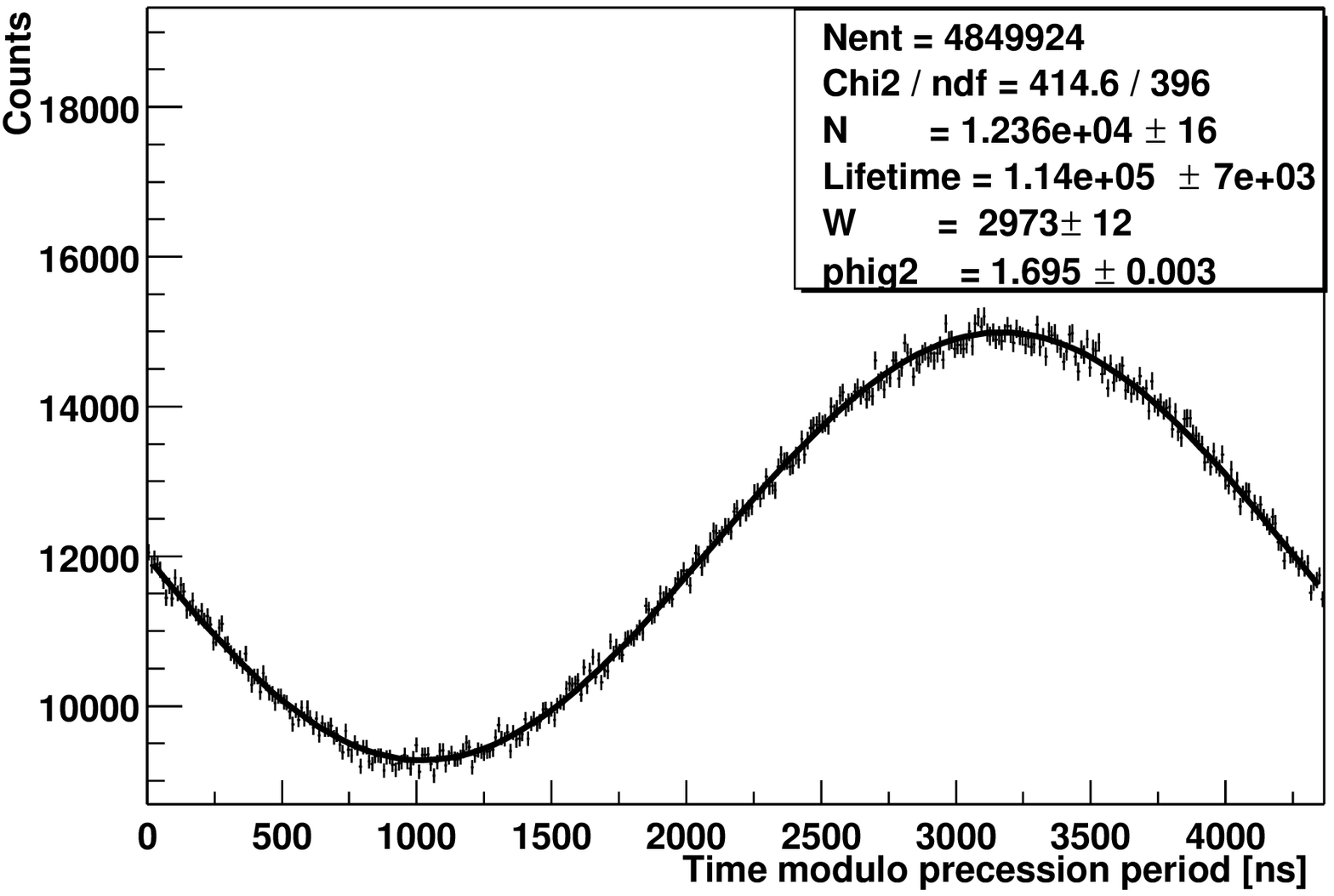} \hfill\\
\includegraphics[width=0.49\textwidth,angle=0]{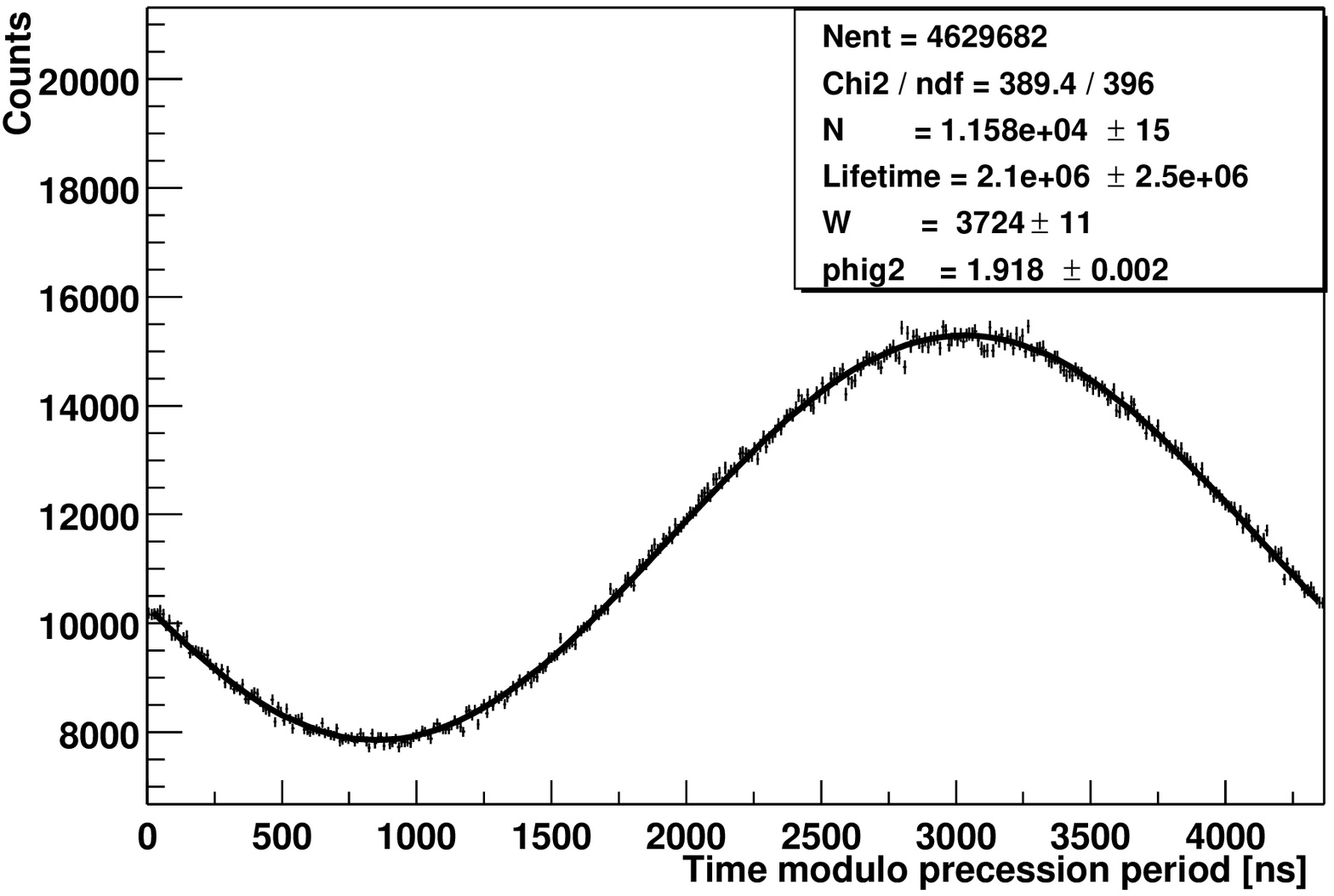} \hfill
\caption{Traceback analysis:
Plots of the data and fits to Eq. ~(\ref{eq:par5}),
modulo the anomalous
precession period, for each running period, as recorded by the traceback
system. (1999 above, 2000 below). The $g-2$ phase parameter (labeled phig2
in the fit box) is used in subsequent vertical angle fits.}
\label{fig:g2redin}
\end{center}
\end{figure}

As indicated above, a non-zero EDM would generate an oscillation in the
vertical angle of decay
positrons, $90^\circ$ out of phase with the $g$-2 number modulation, $N(t)$.
While there are several mechanisms which might generate oscillations in the
measured vertical angle, all are in phase with the number ocillation, 
$N(t)$, and this is also measured in the traceback system. 
By  fixing the relative phase of the EDM and $N(t)$
oscillations, the false EDM signals produced by these effects can be minimized. 


\par
The $N(t)$ spectrum is fit first.
To minimize the effect of periodic disturbances at other frequencies, which
de-phase and average away when many time bins are combined, the data are plotted
vs. time, modulo the $g$-2 precession period.  The spectrum is fit to 
Eq. ~(\ref{eq:twcpar5}) where 
the precession period $T=2\pi/\omega=4365.4$~ns, is
fixed by the result of the anomalous precession  analysis.
\begin{equation}
        N(t) = e^{-t/\tau_e}(N_0 + W\cos(\omega t + \Phi)),
\label{eq:twcpar5} 
\end{equation}
where $\tau_e$ is an empirical term which parameterizes both 
the effect of muon 
decay and that of the recovery of the chambers, which are disabled during
injection.  The phase is determined to better than 3~mrad~(see
Fig.~\ref{fig:g2redin}).  

Once the precession phase is established, a plot of average
vertical angle versus time,
modulo the $g-2$ precession period,  
is fit to the function
\begin{equation}
        \theta(t) = M + A_{\mu} \cos(\omega t + \Phi) + A_{\text{EDM}} \sin(\omega t + \Phi).
\label{eq:TWCfit}
\end{equation}
where $\omega$ remains fixed as before and $\Phi$ is set by the previous fit. 
The amplitude of the sine 
term represents the EDM signal.

\begin{figure}[h] 
\begin{center}
\includegraphics[width=0.49\textwidth,angle=0]{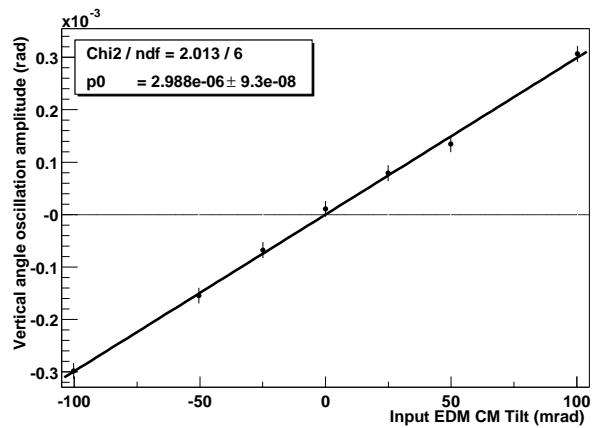} \hfill
\caption{Traceback EDM fit amplitude vs. simulated input precession
plane tilts.  The fit function is a strict proportionality: the slope 
is parameter $p_0$ with no
constant term.  A 
vertical oscillation with amplitude $3\times 10^{-6}$ radians 
corresponds to 1 $\times 10^{-3}$~rad of precession plane tilt.}
\label{fig:simedmrelation00} 
\end{center} 
\end{figure}

The measurement of the vertical angle in the lab frame must be converted to a
precession plane tilt angle in the muon rest frame.  (See Eq.  
~(\ref{eq:tipping})).
The conversion factor is
determined through simulation.  Several sets of simulated trajectories were
generated and reconstructed, each having a different value for precession plane
tilt.  Shown in Fig.~\ref{fig:simedmrelation00} is the fit EDM amplitude for
various input precession plane tilts.  A
3~$\mu$rad amplitude vertical oscillation is generated for each milliradian of
precession plane tilt.

\begin{figure}[h] 
\begin{center}
\includegraphics[width=0.49\textwidth,angle=0]{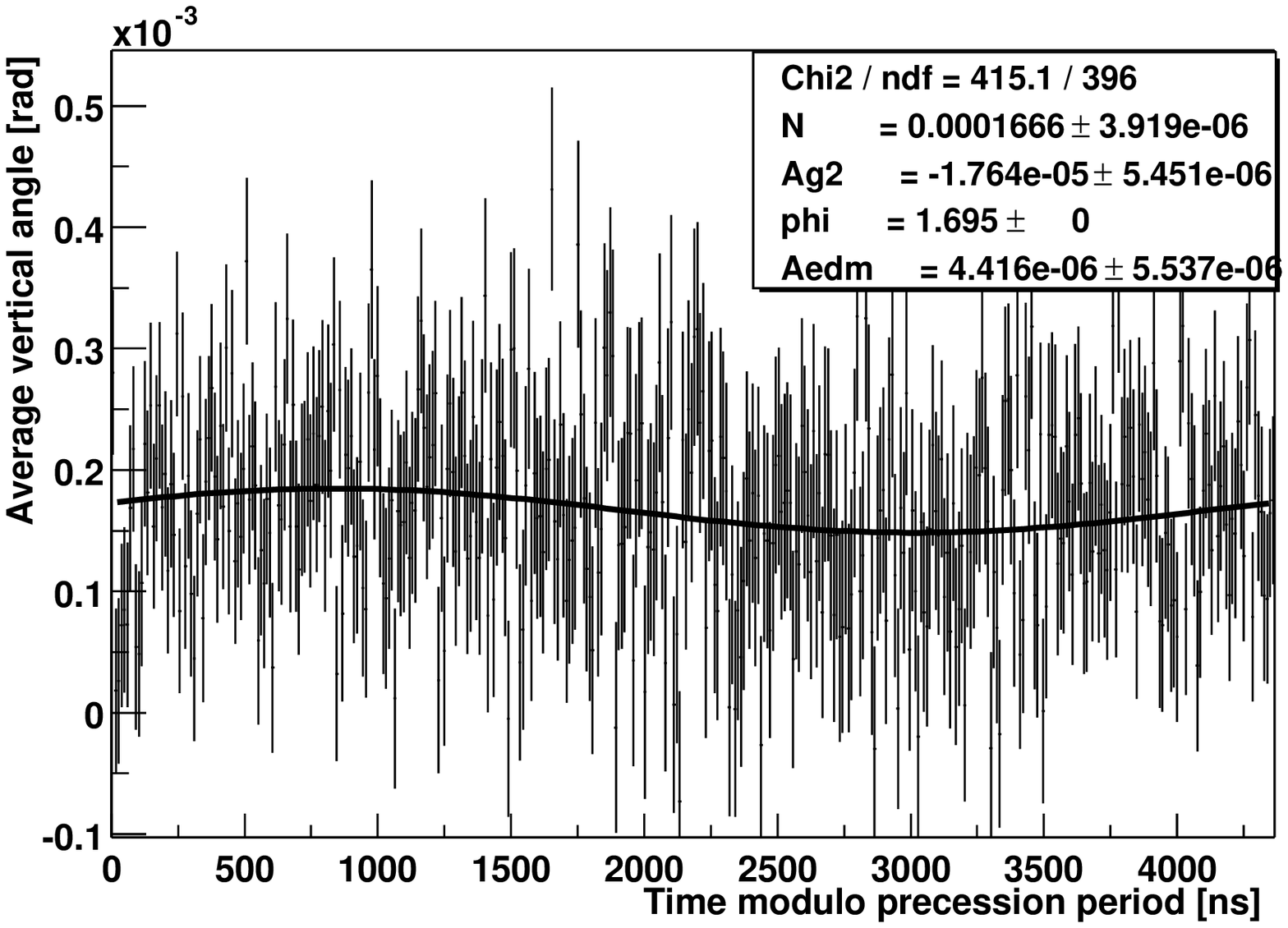} \hfill
\includegraphics[width=0.49\textwidth,angle=0]{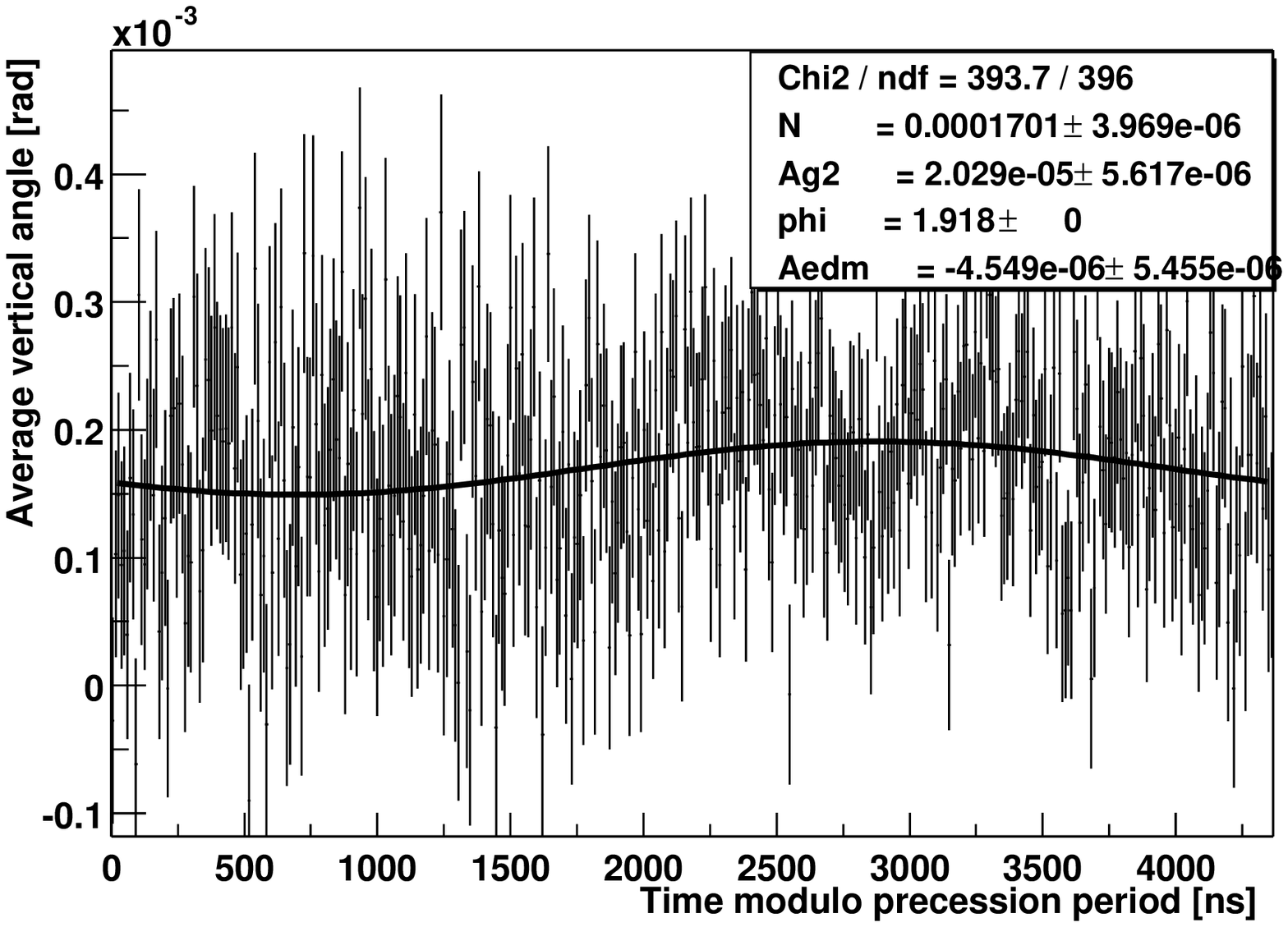} \hfill
\caption{Traceback analysis: Average vertical angle vs. time,
modulo the anomalous precession period, measured by the
traceback system for each running period (1999 above, 2000 below).
The data are fit to Eq. ~(\ref{eq:TWCfit}). The amplitude of the oscillations, 
$A_{{\text {EDM}}}$, is proportional to the EDM. }
\label{fig:edmredin}
\end{center}
\end{figure}

The 1999 and 2000 data-sets contain, respectively, approximately 4.8 million
and 4.6 million well-fit tracks.  In each data set, approximately 15\% of the
tracks are background: mis-constructed tracks or tracks
scattered from the upstream vacuum chamber or 
previous calorimeter
stations. To
reduce the level of mis-constructed tracks to the percent level
requires cuts which would throw away most of the data. The more liberal cuts
chosen for the final data sample do not induce a false EDM signal.
The time spectrum of the average vertical angle
combined into one histogram with
a length of the precession period, is then fit with the sum of a sine and 
cosine,
as described above.  Fits for each data-set
are shown in Figs.~\ref{fig:edmredin}.  
The results are 
$(4.4~\pm~6.3)\times 10^{-6}$~rad oscillation for the 1999 data-set and 
$(-4.5~\pm~6.2)\times 10^{-6}$~rad for the 2000 data-set.  
Combining the results, an
oscillation amplitude of $(-0.1~\pm~4.4)\times 10^{-6}$~rad is obtained,
where the error is statistical. 

\begin{table*}[t]
\caption{Table of systematic errors from the traceback analysis.}
\begin{center}
\begin{tabular*}{0.5\textwidth}{|p{0.15\textwidth}|p{0.08\textwidth}|p{0.08\textwidth}|p{0.1\textwidth}|}
\cline{1-4}
& & & \\
Systematic error & Vertical oscillation amplitude ($\mu$rad~lab) & Precession plane tilt (mrad) & False EDM generated
10$^{-19}$ ($e\cdot$~cm) \\ 
\cline{1-4}
Radial field & 0.13 & 0.04 & 0.045 \\
Acceptance coupling & 0.3 & 0.09 & 0.1 \\
Horizontal CBO & 0.3 & 0.09 & 0.1 \\
Number oscillation phase fit & 0.01 & 0.003 & 0.0034 \\
Precession period & 0.01 & 0.003 & 0.0034 \\
Totals & 0.44  & 0.13 & 0.14 \\
\cline{1-4}
\end{tabular*}
\end{center}
\label{tab:tbsyserror}
\end{table*}

Many systematic uncertainties have been studied and those relevant to the
measurement are listed in Table~\ref{tab:tbsyserror}.  The ``radial field''
error refers to the fact that an average radial magnetic field around the ring
would tilt the precession plane in the same way as an EDM.  The effect of 
a radial magnetic field on the decay positrons, which would 
change the vertical angle of 
the tracks, can be neglected. 
Another uncertainty
comes from the geometry of the detector. 
In making their way to the detectors, positrons with an intial outward 
radial momentum component 
travel further,
on average, than those with an initial inward radial momentum component.  
Combined with varying vertical angle
acceptance for different decay azimuths and an off-center muon distribution, a
radial beam oscillation can appear in the detector data as a vertical
oscillation.  An error on the number
oscillation phase or period may result in some mixing between the number
oscillation signal and the precession plane tilt signal.  The total systematic
error is $0.14~\times~10^{-19}~(e\cdot$cm). For more details 
see ~\cite{sossong}. The negligible systematic errors indicate that the
TWC method should be considered in any future attempt to measure the muon EDM.

Since the systematic errors are negligible, 
the traceback system's 
measurement of the EDM for the positive muon is determined by the value and 
(statistical) error for the vertical oscillation amplitude alone: 
$(-0.04\pm 1.6)\times 10^{-19}$~($e\cdot$cm), which corresponds to an  upper 
limit of

\begin{equation}
|d_{\mu^+}| < 3.2~\times~10^{-19}~(e\cdot\text {cm})~(95\% ~\text{C.L.}).  
\end{equation}

\section{FSD Analysis of the year 2000 $\mu^+$ data set}\label{sec:ronsanalysis}

\begin{figure}[h] 
\begin{center}
\includegraphics[width=0.49\textwidth,angle=0]{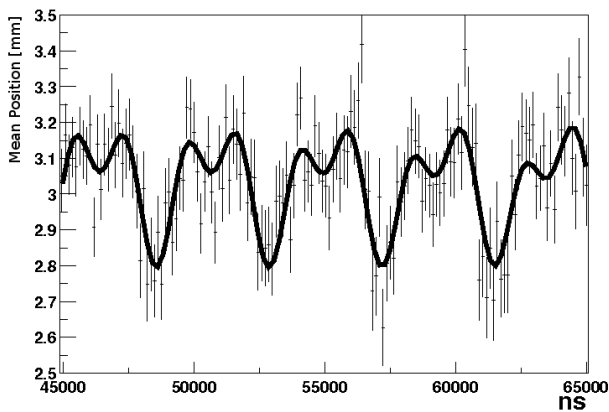} \hfill
\caption{Sample plot of the mean position 
on an FSD versus time in fill with
fit to Eq. ~(\ref{eq:meanfit}) overlaid.  
Oscillations at the $g-2$~($T \approx 4365 $~ns) and CBO~($T \approx 2150$ ~ns) 
frequencies are visible. The 3~mm vertical offset was later corrected.
(See Figs. ~\ref{fig:svsstatb} and ~\ref{fig:svsstata}, below).}
\end{center}
\label{fig:meanvst}
\end{figure}

As described in the introduction,
a non-zero EDM would result in an oscillation of the mean vertical position at the
$g-2$ frequency but $90^{\circ}$ out of phase with the number oscillation.
Therefore, for each detector, the mean position of hits on the FSD, matched to calorimeter 
events with
energy $1.4-3.2~$GeV, is plotted versus time.  (The center tile is not used in
the mean.)  An example from a single station is shown in 
Fig.~\ref{fig:meanvst}.  The plot is fit to 

\begin{eqnarray}
f(t) &=& K + \left[S_{g2}\sin(\omega t) + C_{g2}\cos(\omega t)\right]  \\
&+& e^{-\frac{t}{\tau_{{\text {CBO}}}}} \times  \left[S_{{\text {CBO}}}\sin\left(\omega_{{\text {CBO}}}(t-t_{0}) + \Phi_{{\text {CBO}}}\right) \right. \nonumber \\ 
 & & \left. ~~~~~~~~~+~C_{{\text {CBO}}}\cos\left(\omega_{{\text {CBO}}}(t-t_{0}) + \Phi_{{\text {CBO}}} \right)\right]  \nonumber \\
&+& Me^{-\frac{t}{\tau_{M}}}. \nonumber
\label{eq:meanfit}
\end{eqnarray}
\noindent
The constant term $K$ characterizes the vertical offset between the beam 
and the detector.  There are sine and cosine terms of frequency 
$\omega$ with the
phase fixed such that the cosine term is aligned with the number oscillation.  
Thus, an EDM signal would appear $90^{\circ}$ out of phase,
in the sine term ($S_{\mathrm g2}$);
$\omega$ is fixed to the frequency measured in the $g-2$ analysis.
$t_0$ is an empirical term, fixed to 100 $\mu$sec.

\par
  In addition, there is a sinusoidal term with the frequency of the coherent 
betatron oscillation(CBO).  Horizontal focusing causes the beam
position near any detector \
to oscillate radially at $\omega_{\text{CBO}}\approx 465$~kHz.  
The oscillation is prominent at early times, but de-phasing causes its 
amplitude to decay with a lifetime $\tau_{\text CBO} \approx 100~\mu s$.
The CBO is clearly evident in a plot of the vertical profile 
width versus time. 
The frequency $\omega_{\mathrm CBO}$, lifetime $\tau_{\text CBO}$, 
and phase
($\Phi_{\text CBO}$) of the CBO are obtained from fits to the width. 
Those parameters are then fixed 
in the fit to the average 
position versus time, Eq. ~\ref{eq:meanfit}
The final term, with $\tau_{\text M}$ fixed to $60~\mu$s, 
accounts for slow changes in detector response and possible pulse pileup 
effects, when two pulses arrive within the detector deadtime.

\begin{figure}[h] 
\begin{center}
\includegraphics[width=0.49\textwidth,angle=0]{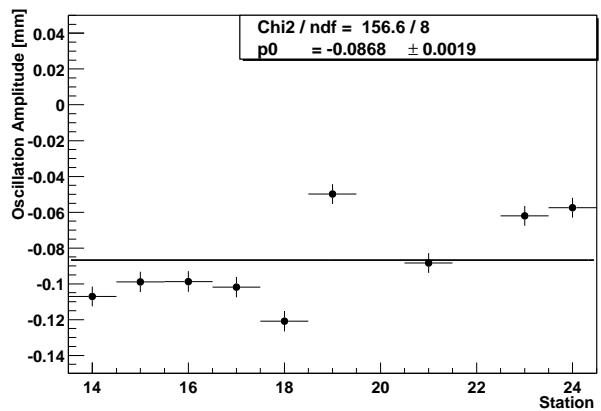} \hfill
\caption{FSD analysis: $S_{\text g2}$ versus station before the beam 
realignment in the year 2000 data period. The offset from zero, 
indicated by fit parameter $p_0$, and 
inconsistencies between detectors indicated by
the large $\chi^2$ of the fit,
are due to 
the misalignments between the detectors and the beam.}
\label{fig:svsstatb}
\end{center}
\end{figure}

\begin{figure}[h] 
\begin{center}
\includegraphics[width=0.49\textwidth,angle=0]{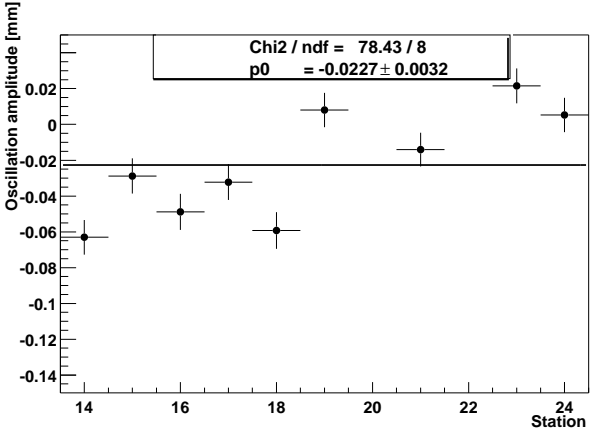} \hfill
\caption{FSD analysis:
$S_{\mathrm g2}$ versus station after the beam realignment.
The overall offset from zero is smaller when the beam is better aligned but 
the inconsistencies between detectors, indicated by a $\chi^2$ which is
only somewhat improved, remain. The line represents a best fit constant
to the data.
}
\label{fig:svsstata}
\end{center}
\end{figure}

Approximately two-thirds of the way through the year 2000 data run, the stored 
beam was moved downward by 2~mm to improve its alignment with the detectors.  
Fig.~\ref{fig:svsstatb} shows the fitted 
value of the EDM term ($S_{\mathrm g2}$) 
versus station number
before the alignment. The figure shows significant variations between 
detectors with an
average of 0.087~mm, corresponding to an EDM nearly as large as the CERN
limit.  Fig.~\ref{fig:svsstata} is a similar plot from data taken
after the beam alignment.  
Although the amplitudes of oscillation are much reduced after beam 
alignment, there are still unacceptably large variations among detectors.
\par
As explained above, these inconsistencies arise from residual
misalignment between the detectors and stored beam combined with oscillations 
in the width of the positron vertical profile at the $g-2$ frequency.  
While data taken after the beam was realigned show a significantly
smaller oscillation, if the 
alignment were perfect and the EDM zero, there would be no oscillation at all. 
The amplitude of an oscillation caused by an EDM should be consistent from
detector to detector.
\par
The CBO oscillation that was described earlier can be used to eliminate the 
large false EDM signal that results from detector misalignment. The CBO
causes an oscillation in the width of the vertical profile because, for
example, decay  positrons must travel further to the detectors when the 
muons are at their maximum radial position. The oscillation in the mean 
vertical position at the CBO frequency noted in Fig. ~\ref{fig:meanvst}
results from the
misalignment of the detector with the plane of the beam. Since the 
vertical oscillation at the CBO frequency is sensitive to the detector 
misalignment, it can be used to correct for the corresponding systematic error
in the EDM measurement.

\begin{figure}[h] 
\begin{center}
\includegraphics[width=0.49\textwidth,angle=0]{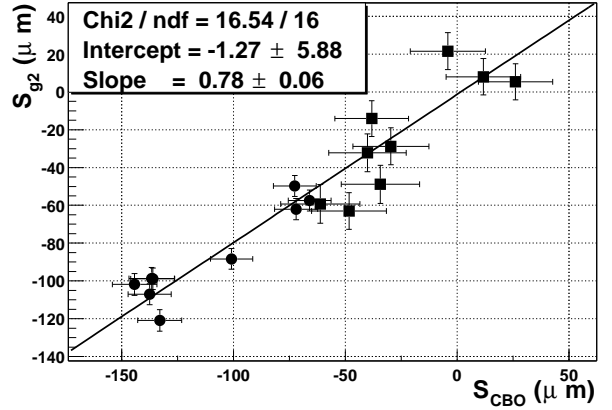} \hfill
\caption{FSD analysis: the
$g-2$ sine amplitude vs. CBO sine amplitude from fits to the 
mean vs. time.
}
\label{fig:g2vscbo}
\end{center}
\end{figure}

A plot
of the vertical oscillation amplitude at the 
$g-2$ frequency ($S_{\mathrm g2}$)
versus the vertical oscillation amplitude 
at the CBO frequency ($S_{\text {CBO}}$) is 
shown in Fig.~\ref{fig:g2vscbo}.  The 18 points shown are data from
the 9 FSDs used in the analysis before and after beam realignment.
The fit to a line produces a good
$\chi^{2}$, showing, as indicated earlier, that the amplitudes of the 
two oscillations are well
correlated.  The $y$-intercept, where the  CBO oscillation disappears,
corresponds to an EDM oscillation measurement made by a
perfectly aligned detector, which thus 
eliminates a large systematic error.

From simulation, the expected vertical oscillation due to an EDM is $8.8 ~\mu$m
per $10^{-19}~(e\cdot$cm).  The $S_{g2}$ intercept, 
\begin{equation}
S_{g2}(0) = (1.27 \pm 5.88) ~\mu{\text m},
\label{eq:sg2}
\end{equation}
implies an EDM measurement of 
\begin{equation}
|d_{\mu^+}|<(0.14 \pm 0.67)\times 10^{-19}~(e\cdot {\text {cm}}),
\label{eq:sg2edm}
\end{equation}
using only the statistical error. A nearly identical relation between 
oscillation amplitude and EDM limit is
obtained in the PSD analysis, presented below.
\par
Systematic errors dominate the measurement.  Any
source of vertical oscillation at either the $g-2$ or CBO
frequency, in the correct phase, is a potential source of systematic error.  
The largest of the errors is due to the tilt of the detectors.  A tilt
in the detector around the beam direction
combined with a horizontal oscillation in the impact position on the 
detector face, would result
in an apparent vertical oscillation.  Simulation studies of decay positron
tracks reveal no horizontal
oscillation on the detector face at the $g-2$ frequency in the EDM phase, but
it does indicate a horizontal oscillation at the CBO frequency of amplitude 
0.6~mm.  
Measurements with a level established that the 
average detector tilt was less than 8.7~mrad  ($\frac{1}{2}^{\circ}$), 
implying a systematic error of 
$6.1~\mu{\mathrm m}$ on $S_{g2}$. A more direct 
measurement of detector tilt, which arrived at a much more
stringent limit, was made 
with the PSD. (See section ~\ref{sec:petersanalysis}, below).
\par

Another systematic error could arise if the electrostatic
quadrupoles were themselves tilted with respect to the storage
ring field. In this case, a small component of the nominally radial CBO
oscillation frequency would be manifest  as an oscillation
in the vertical mean position of the decay positrons, even in a
perfectly aligned detector. Surveys of the quadrupoles indicate a tilt
of less than 2 mrad; the resulting systematic error is 3.9~$\mu$m.

If the average vertical spin component of the muon is not zero then neither
is the average vertical angle of decay positrons.
Since there is a longer average path length for positrons emitted when 
the muon spin is outward than inward, this would result in an apparent vertical
oscillation at the detector face, in phase with the expected EDM signal.  The 
average vertical angle of positrons 
approaching a detector was measured by the traceback detector.  The angle 
was combined with the change in path length for positrons from decays when 
the muon is pointed inward versus outward, obtained from simulation, to
give an estimate of the vertical oscillation at the $g-2$ frequency due to muon
vertical spin.
This effect is also present at the CBO frequency since positrons from decays
when the muons are further out have longer path lengths to the detector.  
This leads
to a partial cancellation of the systematic error, which is $5.1~\mu$m,
overall.  

  A radial magnetic field would deflect decay positrons vertically, causing 
a similar systematic error.  Although the 
{\it average} radial field in the ring is known to be less than 20
parts per million (ppm) 
because it 
affects the beam position, the local radial magnetic field felt by the 
decay positrons may be larger than 100~ppm.  
The size of the error can be 
estimated in an analysis similar to that used for the muon vertical spin.  
Once again, the error of $2.5~\mu$m reflects a partial cancellation from the
CBO. It should be noted that the average part of the radial magnetic 
field affects the spin similarly to an EDM, thus changing the $\Psi$ in
Eq. ~(\ref{eq:par5pm}). However, the radial field, which is only 
20 ppm of the main vertical field, affects the spin at a level which is
two orders of magnitude smaller than our experimental sensitivity to an
EDM. The reason is that in spin dynamics, the average magnetic field effect
is almost completely canceled by the average field of the focusing
electrodes (while in beam dynamics, they cancel each other completely).
And without an average radial magnetic field, the average electric
field cannot appear in the $g-2$ storage ring with its homogeneous vertical
magnetic field.
\par
Since each of the top and bottom of the calorimeters is read out 
by a different PMT,
there is the potential for timing and energy calibration offsets.  
By applying different calibration constants to the top and bottom of
each detector, the relative gains of the two halves were corrected to 
within 0.5$\%$. The residual error results largely from the vertical
offset of the beam during the year 2000 data-taking run. For the 2001 run,
where the offset was much smaller, the corresponding error is about 0.1$\%$.
(See section ~\ref{sec:petersanalysis}).
The timing difference was estimated using the cyclotron
structure of the data seen early in each fill, before the beam de-bunches.
The time at which the muon bunch passes each detector can be seen as a rate
oscillation with a 149.2 ns period.  By measuring the time of the peak of
each bunch in the top and bottom of the calorimeter the average timing 
difference was found to be less than 0.5~ns.
To determine the potential effects of
these asymmetries, analyses were performed with offsets in the timing and
calibration intentionally inserted.  Based on these analyses, systematic errors
due to timing and calibration  offsets were estimated to be 
3.2~$\mu$m and 2.8~$\mu$m respectively.

The sensitivity of the FSD tiles to low energy back-scatter from the 
calorimeters("albedo") may also cause a systematic error.  
Albedo causes multiple tiles to fire at the same time, giving the
appearance of two positrons hitting the detector when there was only one.  
Albedo is not a problem unless there is a difference in the sensitivity to
albedo in the top and bottom of the detector.  A limit on the size of this 
effect was determined by using several methods of dealing with apparent double
hits (counting both, counting neither, counting one at random).  No deviations
in the results were found larger than 2.0~$\mu$m, which is taken to be the 
systematic error.

There are error bars on both abscissa and ordinate in Fig. ~\ref{fig:g2vscbo}.
Our fits, which use an approximation for the latter error,
introduce an additional systematic error of 1~$\mu$m. The
effect of FSD tile inefficiency and dead time was also investigated but the
resulting systematic errors were less than 1~$\mu$m  and so were 
ignored.

\begin{table*}[h]
\caption{FSD analysis: Error Table for $S_{g2}$, the EDM-sensitive
parameter. The conversion to error on the EDM is $0.114\times 10^{19}$ e-cm/$\mu$m.}
\begin{center}
\begin{tabular}{|c|c|} \hline
Effect   & \hspace{1cm} Error ($\mu$m) \hspace{1cm}  \\
\hline \vspace{-3.7mm} \\
\hline \vspace{-3.7mm} \\
\hline
Detector Tilt &  6.1    \\
Vertical Spin   &  5.1     \\
Quadrupole  Tilt   & 3.9          \\
Timing Offset & 3.2       \\
Energy Calibration   & 2.8   \\
Radial Magnetic Field   & 2.5   \\
Albedo and Doubles & 2.0   \\
Fitting Method & 1.0   \\
Total Systematic & 10.4   \\
Statistical & 5.9   \\
\hline
Total Uncertainty &  11.9  \\
\hline
\end{tabular}
\end{center}
\label{tab:sys}
\end{table*}

The systematic effects, given in Table~\ref{tab:sys}, are uncorrelated;
the uncertainties are added in quadrature.  The total error is 11.9 ~$\mu$m,
which gives an EDM measurement of 
\begin{equation}
d_{\mu^+} = (-0.1 \pm 1.4)\times 10^{-19}~(e\cdot {\text {cm}}). 
\end{equation}
and a limit
\begin{equation}
|d_{\mu^+}| < 2.9~\times~10^{-19}~(e\cdot\text {cm})~(95\% ~\text{C.L.}).  
\end{equation}

\section{PSD analysis of 2001 ($\mu^-$) data}\label{sec:petersanalysis}

The Position Sensitive Detector (PSD) \cite{prisca} is a finely segmented
two-dimensional scintillator hodoscope covering the positron entrance faces of
five detector stations. 
Each hodoscope consists of one
plane of 20 scintillator tiles directed horizontally and one plane  
of 32 tiles directed vertically. Each tile is 8 mm wide and 7 mm thick. 
The PSDs were mounted on the face of the calorimeters, providing vertical and
horizontal position data. 
As indicated in the introduction, two EDM searches were made with the PSD,
one modeled on that made with the FSD and another in which the symmetry
of $\Phi$ with $y$ was examined.
\par
While the systematic concerns of the first search are closely related
to those of the FSD,
the latter search requires that
the correct $t=0$  be established for every 
tile of the detector. Using coincident timing information from the
calorimeter, the time offsets for every PSD stick can be determined,
on a run-by-run basis, and the hits in the tiles aligned with the calorimeter
within a 25 ns time window. 
Moreover, 
since the phase is a strong function of the decay positron energy, an elaborate
program of gain balancing was also required. First, the overall
gain  of each calorimeter was established, on a run-by-run basis. To this end, 
the very linear region of the energy distribution, from about 60
to 90\% of the maximum, was fit to a straight line. The x-intercept
was taken as the energy endpoint, 3.1 GeV. To minimize pulse pileup,
only data taken more than 250 $\mu$sec after injection were included
in the calibration energy spectra. 
After energy scale adjustments and corrections for timing offsets,
calorimeter times and energies could be matched with a PSD vertical 
coordinate.
A similar calibration procedure then established {\it position
dependent} energy endpoints. In this case, where because of shower
leakage energy endpoints are not expected to sit at 3.1 GeV, the spectra
were matched to the predictions of a complete tracking and detector simulation.
\par
Time-dependent detector response can be misinterpreted as an EDM signal.
Variations in gain with time were studied separately and
eliminated by applying a 
time-dependent gain factor, $f(t,Y)$ for each $Y$ tile. The correction
function took the form
\begin{equation}
  f(t,Y) = (1 + A_{1}e^{-t/A_{2}})(1 + A_{3}e^{-(t-A_{4})^2/A_{5}^2}),
  \label{eq:gain}
\end{equation}
where parameters $A_1 - A_5$ depend on the PSD vertical coordinate $Y$.
Pulse pileup was corrected, on an average basis, in the time
spectra. ~\cite{PRD}
Finally, time dependence arising from the cyclotron and vertical
betatron motions of the beam were minimized with a simple digital filter.
\par
In order to study systematic errors related to the CBO, in 2001 
the storage ring was run at several different field indices $n$. Also,
approximately two thirds of the data were collected when the average radial
magnetic field was about 10 ppm of the main vertical B field, 
which produced an average vertical
beam displacement of 0.6 mm. After the radial field was zeroed, the alignment 
of the beam with respect to the detectors was much improved. Data taken
before and after the magnetic radial field correction were analyzed separately.
Data were also divided into subsets determined by different run conditions.  
\par
A few cuts were made to the data sample of PSD hits and calorimeter pulses.
For single PSD cluster events (no sign of pileup) the energy was restricted
to fall between 1.4 and 3.4 GeV. Fits to time spectra were started
no earlier than 32 $\mu$sec after injection.
Table~\ref{tab:total_pos_count} 
lists the total number of events (after cuts and
before pileup subtraction) for the PSD analyses.

\begin{table*}
\begin{center}
\caption{Number of decay electrons used for PSD data analysis (in millions).}
\label{tab:total_pos_count}
\begin{tabular}{|c|c|c|c|c|c|c|}
\hline
{Data Set} & 
{PSD 1} & {PSD 2} & {PSD 3} & {PSD 4} & {PSD 5} &
{PSD (5 stations)} \\
\hline 
{Before B Correction n = 0.122} &
{66} & {56} & {61} & {42} & {70} &  
{295} \\ 
{Before B Correction n = 0.142} &
{61} & {54} & {60} & {44} & {71} &  
{290} \\ 
{After B Correction n = 0.122} & 
{83} & {73} & {76} & {65} & {93} & 
{390} \\
\hline  
{Total} &
{185} & {157} & {173} & {139} & {209} &   
{975} \\ 
\hline
\end{tabular}
\end{center}
\end{table*}

\subsection{Details of the two PSD Analyses}
\subsubsection{Phase Method}

\begin{figure}[h]
\includegraphics[width=0.5\textwidth]{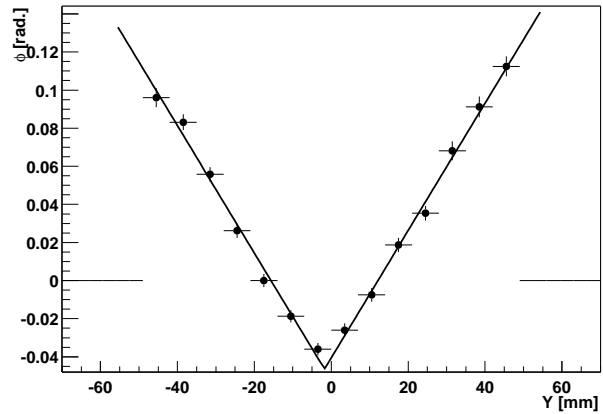}
\caption{PSD analysis:
PSD 15 before B$_r$
change. Phase of $g-2$ oscillation vs. vertical position, with fit to 
Eq. ~(\ref{eq:psd_phase_fit}). $p_1$, the EDM-sensitive parameter, describes
any asymmetry between the two sides of the graph..}
\label{fig:Phase_Fit}
\end{figure}

\begin{figure*}
\subfigure[Before B$_r$ Correction]
{\includegraphics[width=0.49\textwidth]{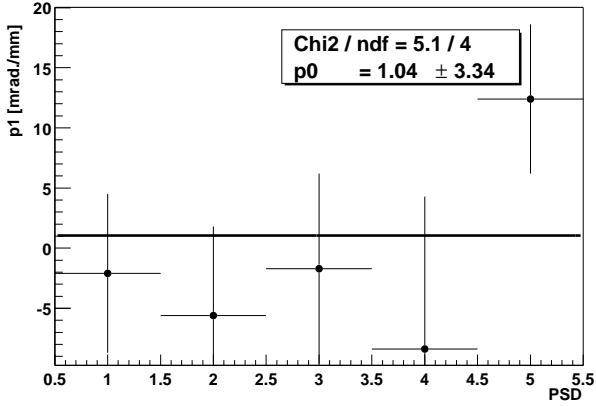}}
\subfigure[After B$_r$ Correction]
{\includegraphics[width=0.49\textwidth]{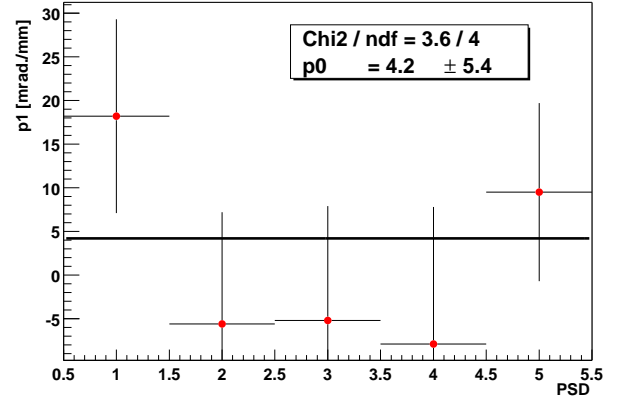}}
\caption{PSD analysis: EDM-sensitive parameter $p_1$ 
(see equation ~\ref{eq:psd_phase_fit})  versus PSD station. Horizontal line
indicates fit to a constant.
$p_1$ is consistent with zero for both the
data taken before and after $B_R$ was changed (see text for discussion).}
\label{fig:Edm_1}
\end{figure*}

The ratio method ~\cite{PRD} was used to determine the
$g-2$ phase vs. vertical coordinate for each of the 20 horizontally-directed 
tiles
on each detector. The three parameter ($A$, $\omega$ and $\Phi$)
fits of the  ratio method are particularly insensitive to any
residual slow variations in detector performance. The phase vs. vertical
coordinate plot was then fit to the 4-parameter function
\begin{equation}
  \Phi(y) = p_o + p_{1}(y -  p_{2}) + |p_{3}(y - p_{2})|.\,
  \label{eq:psd_phase_fit}
\end{equation}
where $p_1$, which is proportional to the EDM, describes the up-down 
asymmetry~(see \cite{giron04TH} for details including the determination
of the proportionality constant used below). Parameter $p_3$ accounts for
phase changes not related to the EDM signal, $p_2$ corresponds to the
vertical detector offset and $p_0$ is an overall detector phase offset.
The result of the fit for one detector, for one data set, is plotted 
in Fig.~\ref{fig:Phase_Fit} and the $p_1$-variable for all PSD fits, 
both before and after the radial magnetic field
correction 
are plotted versus PSD station in Fig.~\ref{fig:Edm_1}. 


\subsubsection{Vertical Position vs. Time Method}
The vertical position vs. time method is modeled closely on that used
in the analysis of the FSD data and serves as a cross-check 
on the phase vs. vertical position
analysis. For each PSD station, the average vertical position and rms
width were fit vs. time. The rms vertical width was fit first in order to
determine the CBO period and lifetime parameters ($T_{{\text {CBO}}}$ and $\tau_{{\text {cbo}}}$), which were
then fixed in the fit of the mean vertical position. The rms vertical 
width distribution
vs. time, was fit to the 9-parameter function:
\begin{eqnarray}
f_{{\text {RMS}}}(t) = a_o + A_{1}\sin\left(\frac{2\pi t}{T}\right) +
B_{1}\cos\left(\frac{2\pi t}{T}\right) \nonumber \\
+ A_{2}\sin\left(\frac{4\pi t}{T}\right) + B_{2}\cos\left(\frac{4\pi~t}{T}\right) \nonumber \\
+ e^{-\frac{t}{\tau_{{\text {cbo}}}}}\left(A_{{\text {cbo}}}\sin\left(\frac{2\pi t}{T_{{\text {cbo}}}}\right)+ B_{{\text {cbo}}}\cos\left(\frac{2\pi t}{T_{{\text {cbo}}}} \right) \right), 
\label{eq:psd_rms_fit}
\end{eqnarray}
where $a_o$ is the average PSD profile width. The time bin-width was set to the
cyclotron period, 149.185~ns, in order to average out oscillations in the time
distribution caused by the bunched structure of the beam in the storage ring.

The parameters $T_{cbo}$,
$\tau_{{\text {cbo}}}$ and $T$, 
the muon spin precession period $(2\pi/\omega)$ are all fixed. In particular, 
$T$ is set
to the nominal value 4365.4~ns. The rms width oscillations contain
precession frequency terms $A_1$ and $B_1$, double precession frequency terms
$A_2$ and $B_2$ due to the changes in average energy, detector acceptance 
and time-of-flight
of the decay  electron within the precession period. CBO frequency terms 
$A_{{\text {cbo}}}$ and $B_{{\text {cbo}}}$ arise because of variation in the decay electron
time of flight with radius.
\par
Next, the vertical mean position versus time is fit to a similar
function:
\begin{eqnarray}
f_{Ave}(t) = y_o + A_{g2}\sin\left(\frac{2\pi t}{T}\right) + B_{g2}~\cos\left(\frac{2\pi t}{T}\right) \nonumber \\
+ A_{2g2}\sin\left(\frac{4\pi t}{T}\right) + B_{2g2}\cos\left(\frac{4\pi t}{T}\right) \nonumber \\
+ e^{-\frac{t}{\tau_{{\text {cbo}}}}} \times  \left[S_{{\text {cbo}}}\sin\left(\frac{2\pi t}
{T_{{\text {cbo}}}} + \Phi_{{\text {cbo}}}\right)\right. \nonumber \\
\left. + C_{{\text {cbo}}}\cos\left(\frac{2\pi t}{T_{{\text {cbo}}}} +
    \Phi_{{\text {cbo}}} \right)\right],
\label{eq:fsd_mean_fit}
\end{eqnarray}
where  $y_o$ describes any misalignment between the beam and
the vertical center of the PSD.  There are precession frequency
sine and cosine terms, $A_{g2}$ and $B_{g2}$, and double 
precession frequency terms $A_{2g2}$ and $B_{2g2}$
all with the $g-2$ period fixed. The absolute phase of each detector was chosen
so that the $g-2$ number oscillation is described by $B_{g2}$ and any EDM
signal would appear in $A_{g2}$. There are also CBO sine and cosine terms,
$A_{{\text {cbo}}}$ and $B_{{\text {cbo}}}$.
The CBO decay time, period and 
phase are fixed by the RMS width fit.
\par
\begin{figure}[h] 
\begin{center}
\includegraphics[width=0.5\textwidth]{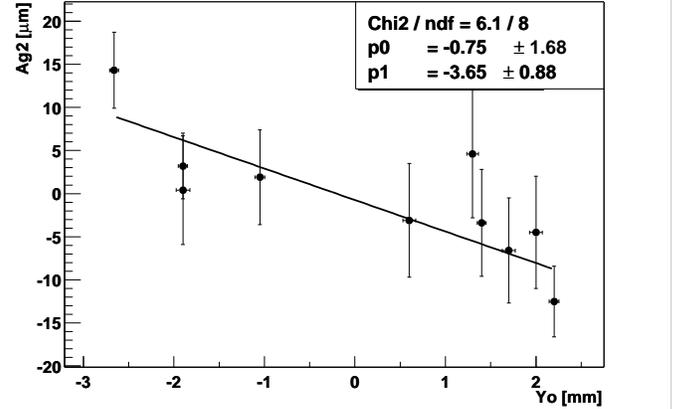} 
\caption{PSD analysis:
Fitted sine wave amplitude, $A_{g2}$  for the mean vertical
position fit versus measured detector offset. A straight-line fit 
is overlaid. Since the intercept
is consistent with 0, $A_{g2}$ is directly proportional to 
the measured detector offset.}
\label{fig:Sin_Yo}
\end{center}
\end{figure}

As in the FSD analysis, the very large false EDM signal
caused by detector misalignment~\cite{mcnabb03TH} must be
corrected. The EDM-sensitive parameter, 
$A_{g2}$, was plotted versus
the PSD measured  vertical detector offset $y_{o}$. The value of
$A_{g2}$ at $y_{o}=0$, obtained 
from a  fit to a straight line, corresponds, once again, to the 
EDM-related amplitude that would
be measured by a perfectly aligned detector.
Fig.~\ref{fig:Sin_Yo}
shows the $A_{g2}$ amplitude for the 5 PSD detectors, with separate points
plotted for the data sets taken before and after the radial field 
correction.

\subsection{Systematic Errors} 
\label{sec:systematics}
Systematic errors must be assessed separately for the two
PSD analyses but most are
similar to those of the FSD analysis. One of the largest systematic
errors for the phase method is associated with the vertical alignment 
of the PSD with respect to the stored beam. To estimate the error,
the relationship  between the size of the false EDM signal and 1 mm of detector
misalignment was established, using Monte Carlo simulation. The
accuracy of the mean vertical position measurement  was 
studied by imposing on the data  a 10 $\%$  tile inefficiency   
and 5$\%$ gain variation for top/bottom halves of the calorimeter. 
The estimated error in vertical position is 0.2 mm. 
The Monte Carlo simulation was then used to determine the size of the
false EDM signal produced per unit misalignment (see \cite {giron04TH}),
140 $\mu$rad/mm. Multiplying that scaling factor by the estimated vertical 
misalignment  yields
an error of 29 $\mu$rad on $p_1$. The resulting error on the EDM is 
$0.74 \times 10^{-19}~e \cdot$cm.
\par
Since the intercept of the sine wave amplitude versus detector offset plot was
used for the vertical position EDM measurement, detector misalignment 
was less important to the overall systematic error.
In that case, the systematic uncertainty was estimated using the
slope of the sine wave $A_{g2}$ versus detector offset $Y_o$ plot combined 
with the error in
the detector offset measurement. The systematic error for the detector
misalignment is 1.7 $\mu$m in $A_{g2}$, resulting in an EDM error of 
$0.2 \times 10^{-19}~e \cdot$cm,
as in Eqs. ~\ref{eq:sg2} and ~\ref{eq:sg2edm}.
\par
If the PSD is tilted with respect to the horizontal plane of the storage ring,
horizontal beam oscillations produce apparent vertical oscillations 
in the average position measured by the PSD. The 
PSD tilt was estimated from a two-dimensional
plot of PSD Y v. X coordinates, for data taken more than 32 $\mu$sec after
injection. The mean vertical position was calculated for each X tile
and the ensemble of means was fitted vs. X to a linear function.
The tilt of five PSDs was consistent with 0 and no larger than 0.75 ~mrad
on average.
\par
PSD tilt could also lead to asymmetric vertical losses which would affect
the phase analysis. Higher energy electrons strike the calorimeter at
larger radius, that is, closer to the storage region. Asymmetric losses
could change the energy spectra and hence the phase of the detected 
electrons vs. vertical coordinate. The size of the error was estimated
with the full Monte Carlo simulation. The induced
EDM signal in parameter $p_1$ per mrad of detector tilt is
26 ~$\mu $rad/mm (see \cite{giron04TH}).
Multiplying that scaling factor by the maximum tilt (0.75
mrad) yields an error 20~$\mu $rad/mm on $p_1$. The associated error on
the EDM is $0.5 \times 10^{-19}~e\cdot$cm. 
\par
As noted before, a time-dependent energy scale difference between
the top and bottom of the calorimeter can induce a false EDM signal.
The energy balance procedure was checked by examining events in which
the signal was restricted to the top or bottom of the calorimeter.
The measured average top/bottom energy scale difference is less than
0.1\% after the time-dependent energy correction was applied.
Monte Carlo simulation indicates 
that a  1 $\%$ top/bottom gain imbalance produces
a 43 $\mu$rad/mm error, corresponding to an error of 4.3 $\mu$rad/mm 
in our case.
\par
In order to study the error due to tile inefficiency, an
artificial 10$\%$ inefficiency for tile 3 (for all PSDs) was imposed
on the data.  (Tile 3 corresponds to the uppermost tile used in
the analysis). The 
intercept change in Fig.~\ref{fig:Sin_Yo} is only 1.8~$\mu$m, 
which is taken as an estimate of the systematic error for the vertical 
mean method. 

\begin{table*}
\caption{Errors for the Phase Method. The first six entries of
the central column list the size of a physical error and the
sensitivity (assumed to be linear) of the EDM result to that error.
The column to the right gives errors on EDM-sensitive parameter
$p_1$ and in the totals, the error on the EDM itself.}
\begin{center}
\label{tab:total_phase_syst}
\vspace{0.2cm}
\begin{tabular}{|l|c|c|}
\hline
{Source} & {Sensitivity} & {Result} \\
\hline 
Detector Tilt & $26 ~\mu$rad/mm/mrad$\times$ 0.75 mrad  & 20 $\mu$ rad/mm \\
Detector Misalignment & 138 $\mu$rad/mm/ mm  $\times$ 0.2 mm & 28 $\mu$ rad/mm\\
Energy Calibration & 43 $\mu$rad/mm/ \% $ \times$ 0.1$\%$ & 4.3 $\mu$ rad/mm\\
Muon Vertical Spin & 1.0 $\mu$rad/mm $\times$ 8$\%$ & 8.0 $\mu$ rad/mm\\
Radial B field & 0.72 $\mu$rad/mm/ppm $\times$ 20.0 ppm & 14.4 $\mu$ rad/mm\\
Timing & 17.0 $\mu$rad/mm/ns $\times$ 0.2 ns & 3.4 $\mu$ rad/mm\\
Total systematic&  & 38  $\mu$rad/mm ($0.93 \times 10^{-19} ~e \cdot$cm )\\
Total statistical &  & 28  $\mu$rad/mm $(0.73 \times 10^{-19} ~e \cdot$cm )\\
\hline 
Total &  & 47  $\mu$rad/mm $(1.2 \times 10^{-19}~e \cdot$cm )\\
\hline
\end{tabular}
\end{center}
\end{table*}

\begin{table*}
\caption{Errors for the Mean Position Method. The first five entries of
the central column list the size of a physical error and the
sensitivity (assumed to be linear) on the residual value of 
the EDM-sensitive parameter
$A_{g2}$ for a nomiminal detector offset of 0. In the last
three rows, the systematic, statistical and total errors on the EDM are
also given.}
\begin{center}
\label{tab:total_mean_sys}
\vspace{0.2cm}
\renewcommand{\arraystretch}{1.2}
\begin{tabular}{|l|c|c|}
\hline
{Source} & {Sensitivity} & {Result} \\
\hline 
Detector Misal. & 4.1 $\mu$m/mm $\times$ 0.4  mm  & 1.6 $\mu$m \\
Detector Tilt & 3.0 $\mu$m/mrad $\times$ 0.75  mrad  & 2.3 $\mu$m \\
Energy Calibration & 28.0 $\mu$m/ $\%  \times$ 0.1 $\%$ & 2.8 $\mu$m \\
Muon Vertical Spin & 0.4 $\mu$m/$\%$ $\times$ 8$\%$ & 3.2 $\mu$m \\
Radial B field & 0.13 $\mu$m/ppm $\times$ 20 ppm  & 2.6 $\mu$m \\
Timing & &  1.5 $\mu$m \\
Tile ineff. &  & 1.7 $\mu$m \\
Fitting & & 1.0 $\mu$m \\
Total Systematic &  & 6.2  $\mu$m ($0.7 \times 10^{-19}~e \cdot$cm ) \\
Total Statistical &  & 1.7 $\mu$m ($0.2 \times 10^{-19}~e \cdot$cm) \\
\hline 
Total &  & 6.4  $\mu$m ($0.73 \times 10^{-19}~e \cdot$cm) \\
\hline
\end{tabular}
\end{center}
\end{table*}

\par
Errors for the phase method are summarized 
in Table~\ref{tab:total_phase_syst}.
The total uncertainty is 50 $\mu$rad/mm on $p_1$,
$1.3 \times 10^{-19}~~e \cdot$cm on the EDM.
Systematic uncertainties for the mean vertical position vs. time method
are presented in Table~\ref{tab:total_mean_sys}.
The total systematic uncertainty for the vertical position method is 9.6 $\mu$m
(or $1.1 \times 10^{-19}~~e \cdot$cm) and statistical error is 
1.7 $\mu$m (or $0.2
\times 10^{-19}~~e \cdot$cm).

\subsection{PSD Summary}
\label{sec:fits_summary}
Results from the two PSD analysis
methods are consistent with zero and with each other. 
Averaged
over the five PSD stations, the EDM signal from the phase analysis 
is $d_{\mu^-} = (-0.48 \pm 1.3) \times 10^{-19}~e\cdot$cm, where  $0.73 \times
10^{-19}e \cdot$cm is the 
statistical error and $1.09 \times 10^{-19}~e\cdot$cm
is the systematic error.
Averaged over the five PSDs, fitting the PSD vertical
position versus time yields an EDM value of
$d_{\mu^-} = (-0.1\pm 0.73) \times 10^{-19} ~e\cdot$cm, where  $0.28 \times
10^{-19} ~e\cdot$cm and $0.70 \times 10^{-19}~e\cdot$cm are the statistical
and systematic errors, respectively. 

\section{Combination of EDM results}\label{sec:combination}

The traceback data set is entirely independent of those used for
the FSD and PSD analyses and there are minimal correlations between the 
systematic errors of the traceback analysis and those of the other two.  

The combined measurement from the traceback and FSD analyses
on the $\mu^+$ is
$d_{{\mu}^+}=(-0.1~\pm~1.0)\times 10^{-19}~e\cdot $ cm, 
providing a limit on the permanent electric dipole moment of the positive muon
of

\begin{equation}  
|d_{{\mu}^+}| \leq 2.1 \times 10^{-19}~e\cdot \textrm{cm}\ (95\% ~{\text {C.L.}}).
\end{equation}
\noindent
The $\mu^-$ result is taken from the PSD mean vertical position analysis:
$d_{{\mu}^-}=(-0.1~\pm~0.7)\times 10^{-19}~e\cdot $cm with corresponding limit 
\begin{equation}  
|d_{{\mu}^-}| \leq 1.5 \times 10^{-19}~e\cdot \textrm{cm}\ (95\% ~{\text {C.L.}}).
\end{equation} 
\par
It is important to note that the errors on the traceback measurement are
entirely statistical while those of the FSD and PSD measurements are dominantly
systematic. Although the increased vertical segmentation of the BNL detectors
is a distinct improvement over the two-paddle CERN III approach, significant 
further
progress with this basic technique would be exceedingly difficult. By contrast,
relatively simple remedies~-~increasing the geometric acceptance and reducing
or eliminating the recovery period after injection~-~would lead to direct
improvements in the traceback result.
\par
In determining joint results from the $\mu^+$ and $\mu^-$ data sets,
we must consider the correlation of errors in the FSD($\mu^+$)  
and PSD($\mu^-$) analyses.
While the FSD and PSD data sets are entirely independent, 
three of the systematic errors: 
the vertical spin, top/bottom calibration and radial B-field
errors are correlated and, although it is impossible to judge the extent of the
correlation, one might assume that the B- and E-fields 
of all bending and focusing 
elements are exactly reversed when changing from positive to negative beam 
and that these three errors are fully correlated. 
\par
The EDMs of the $\mu^+$ and $\mu^-$ are found to be
in accord with the $CPT$ requirement that 
$d_{{\mu}^+}=-d_{{\mu}^-}$, or, in terms of directly measured quantities,
$\eta_{{\mu}^+}=\eta_{{\mu}^-}$. The result is
\begin{equation}
d_{{\mu}^-}+d_{{\mu}^+} = (-0.2 \pm 1.3) \times 10^{-19}{\text ~e}\cdot{\text cm}. 
\end{equation}
\noindent
In summing the results for $\mu^+$ and $\mu^-$ to construct this 
difference,
the correlated systematic
errors in the FSD and PSD analyses should tend to cancel,
as described above. However, because of
uncertainties in the extent of the correlation, as a conservative 
measure, we take them
to be completely uncorrelated.
\par
Under the assumption of $CPT$ invariance, we take a weighted average of the 
$\mu^-$ result and the opposite of the $\mu^+$ results. In this case, 
once again as a conservative measure, we take
the three correlated errors of the FSD and PSD analyses to contribute 
maximally to the total error. All other errors are added in quadrature.
We obtain
\begin{equation}
d_{\mu}  =(-0.1\pm 0.9) \times 10^{-19}~e\cdot {\text {cm}}.
\label{eq:finallimit}
\end{equation}
\noindent
This corresponds to the limit

\begin{equation}  
|d_{\mu}|  \leq 1.9 \times 10^{-19}~ e \cdot \textrm{cm}\ (95\%~{\text {C.L.}}),
\label{eq:bnllimit}
\end{equation} 

\noindent approximately a factor of 5 improvement over the previous limit.
A summary of the results from each detector system,
as well those from CERN III, is shown in 
Table ~\ref{tab:results_summary}. 

\begin{table*}
\caption{Summary of EDM results from the CERN III experiment and from
the three analyses of E821. The units are $\times 10^{-19} e \cdot$ cm in all
cases.}
\begin{center}
\label{tab:results_summary}
\vspace{0.2cm}
\renewcommand{\arraystretch}{1.2}
\begin{tabular}{|l|c|c|c|c|}
\hline 
{Analysis} & {Mean value} & {Stat. Error} & {Syst. Error}
& {Total Error } \\
\hline
CERN (1978)         &     &     &     &     \\
\hline 
($\mu^+$)           & 8.6 & 4.0 & 2.0 & 4.5  \\
($\mu^-$)           & 0.8 & 3.8 & 2.0 & 4.3  \\
Average             & -3.7 & 2.8  & 2.0  & 3.4 \\
\hline
E821                 &     &     &     &     \\
\hline
Traceback ($\mu^+$) & -0.04 & 1.6 & 0.14 & 1.6  \\
FSD ($\mu^+$)       & -0.1  & 0.67 & 1.2 & 1.4  \\
\hline
Total $\mu^+$       & -0.1  & 0.6 & 0.8 & 1.0 \\
\hline 
PSD ($\mu^-$)       & -0.1 & 0.28  & 0.7 & 0.73 \\
\hline  
$(d_{\mu}) $ & -0.1 & 0.2 & 0.9 & 0.9 \\
\hline
\end{tabular}
\end{center}
\end{table*}

%

The observed difference between the standard model and the experimentally
determined values of the muon anomaly, 
$\delta a_{\mu} = 295(88)\times 10^{-11}$~\cite{MdRR2007}, could be
attributed, in 
principle, to a shift in the muon spin precession frequency caused by a 
non-zero EDM. See Eq.~(\ref{eq:biggerwa}). This would lead to an EDM of 
$d_{\delta a_{\mu}} = \pm 2.39(0.36) \times 10^{-19} ~e \cdot$cm.
The new limit on the EDM
(Eq. ~\ref{eq:finallimit})
would predict 
$\delta a_{\mu}({\text {EDM}}) = 0.5(+0.51/-0.33)\times 10^{-11}$. 
The probability that $d_{\delta a_{\mu}}$ is compatible with the limit from
Eq. ~\ref{eq:finallimit} is 2\%, suggesting it is unlikely 
that $\delta a_{\mu}$ arises from 
a non-zero EDM.


\begin{acknowledgments}
We thank  the BNL management, along with  the staff of the
BNL AGS for the strong support they have given the muon $(g-2)$ experiment
over a many-year period.
This work was supported in part by the U.S. Department of Energy,
the U.S. National Science Foundation, the German Bundesminister
f\"{u}r Bildung und Forschung, the Alexander von Humboldt Foundation,
the Russian Ministry of Science,
and the U.S.-Japan Agreement in High Energy Physics.
\end{acknowledgments}

\appendix


\end{document}